\documentclass[fleqn,usenatbib]{mnras}

\usepackage{newtxtext,newtxmath}
\usepackage[T1]{fontenc}
\usepackage{subcaption}
\DeclareRobustCommand{\VAN}[3]{#2}
\let\VANthebibliography\thebibliography
\def\thebibliography{\DeclareRobustCommand{\VAN}[3]{##3}\VANthebibliography}

\usepackage[normalem]{ulem}
\usepackage{graphicx}	% Including figure files
\usepackage{amsmath}	
\usepackage{soul}
\title[Hyperonic Uncertainties in NS, BNS and SN]{Hyperonic Uncertainties in Neutron Stars, Mergers and Supernovae}

\author[H. Kochankovski et al.]{
Hristijan Kochankovski,$^{1,2}$\thanks{E-mail: hriskoch@fqa.ub.edu}
Angels Ramos,$^{1}$\thanks{E-mail: ramos@fqa.ub.edu}
Laura Tolos$^{3,4,5}$\thanks{E-mail: tolos@ice.csic.es}
\\
$^{1}$Departament de F\'{\i}sica Qu\`antica i Astrof\'{\i}sica and Institut de Ci\`encies del Cosmos, Universitat de Barcelona, Mart\'i i Franqu\`es 1, 08028, Barcelona, Spain\\
$^{2}$Faculty of Natural Sciences and Mathematics-Skopje, Ss. Cyril and Methodius University in Skopje, Arhimedova, 1000 Skopje, North Macedonia \\
$^{3}$Institute of Space Sciences (ICE, CSIC), Campus UAB,  Carrer de Can Magrans, 08193 Barcelona, Spain\\
$^{4}$Institut d'Estudis Espacials de Catalunya (IEEC), 08034 Barcelona, Spain\\
$^{5}$Frankfurt Institute for Advanced Studies, Ruth-Moufang-Str. 1, 60438 Frankfurt am Main, Germany
}

\date{Accepted XXX. Received YYY; in original form ZZZ}

\pubyear{2015}

\begin{document}
\label{firstpage}
\pagerange{\pageref{firstpage}--\pageref{lastpage}}
\maketitle

\begin{abstract}

In this work we delve into the temperature-dependent Equation of State (EoS) of baryonic matter within the framework of the FSU2H$^*$ hadronic model, which comprehensively incorporates hyperons and is suitable for relativistic simulations of neutron star mergers and supernovae. To assess the impact of the uncertainties in the hyperonic sector on astrophysical observables, we introduce two additional models, namely FSU2H$^*$L and FSU2H$^*$U. These models cover the entire spectrum of variability of hyperonic potentials, as derived from experimental data.
Our investigations reveal that these uncertainties extend their influence not only to the relative abundances of various particle species but also to the EoS itself and, consequently, have an impact on the global properties of both cold and hot neutron stars. Notably, their effects become more pronounced at large temperatures, owing to the increased presence of hyperons. These findings have direct implications for the outcomes of relativistic simulations of neutron star mergers and supernovae, emphasizing the need of accounting for hyperonic uncertainties to ensure the accuracy and reliability of such simulations in astrophysical contexts.

\end{abstract}


%%%%%%%%%%%%%%%%%%%%%%%%%%%%%%%%%%%%%%%%%%%%%%%%%%

%%%%%%%%%%%%%%%%% BODY OF PAPER %%%%%%%%%%%%%%%%%%

\section{Introduction}\label{sec1}

Neutron stars (NSs) are natural laboratories for testing different models of nuclear matter under extreme conditions.
The lack of terrestrial experimental data for matter at densities larger than $\sim  2-3\rho_0$, with $\rho_0$ being the saturation density at the center of nuclei, has motivated the development of theoretical predictions for the Equation of State (EoS) of such extreme matter. Since the pioneering work of \cite{1960SvA.....4..187A}, many of the theoretical models include exotic hadrons such as hyperons in the NS core, which can reach densities of the order of several times $\rho_0$
(see \cite{Chatterjee:2015pua,FiorellaBurgio2018NuclearSupernovae,Tolos:2020aln,Logoteta:2021iuy,Burgio:2021vgk,MUSES:2023hyz} and references therein).

In spite of the lack of experimental constraints for large densities, there exist several recent astrophysical constraints that have to be fulfilled. First, the maximum mass that the model predicts has to be larger than about $2M_{\odot}$ \cite{Demorest2010ShapiroStar,Antoniadis:2013pzd,Fonseca2016,Cromartie2020RelativisticPulsar,Romani:2022jhd}. Second, the dimensionless tidal deformability for a system with chirp mass $\cal{M}$ $= 1.186 M_{\odot}$ has to be about  $\tilde{\Lambda} = 300_{-230}^{+420}$ \cite{LIGOScientific:2018hze}, which is a tighter constraint than the upper bound $\tilde{\Lambda} \lesssim 800$ established by the first analysis of the GW170817 event \cite{LIGOScientific:2017vwq}.
 And, third, the model should reproduce the NICER measurements of a radius of $R \sim 12-14 $ km for a NS of $M \sim 1.3-1.4 M_{\odot}$ \cite{Riley:2019yda,Miller:2019cac} and, similarly, for $M \sim 2 M_{\odot}$ \cite{,Miller:2021qha,Riley:2021pdl}. The aforementioned astrophysical constraints  thus set important restrictions on the intermediate and high density regions of the EoS. In fact, they significantly reduce the number of applicable hyperonic models
for the core of NSs, as the EoS is softened when hyperons appear, leading e.g. to a reduction of the maximum mass the star can sustain.  Additionally, a very stiff nucleonic part of the EoS is ruled out by the radius measurements which favor softer nucleonic interactions.

Furthermore, while a NS can be considered as a cold object, modelling the early stages in the evolution of a NS after a supernova explosion as well as the eventual merging of two NSs requires a finite temperature treatment of the EoS, for which the simple $\Gamma$-law approach, inspired on the thermal behavior of an ideal gas, is usually taken. 
Within this approach, the finite temperature effects enter through the so-called thermal index $\Gamma_{\rm th} = 1 + \frac{P_{\rm th}}{\epsilon_{\rm th}}$, where  $P_{\rm th}=P(T)-P(T=0)$ indicates the thermal pressure and $\epsilon_{\rm th}=\epsilon(T)-\epsilon(T=0)$ is the thermal energy density. A constant value of $\Gamma_{\rm th} \sim 1.75$ is usually considered as it describes the finite-temperature behaviour of the nucleonic EoSs quite reasonably.  
However, 
the necessity of going beyond the usual $\Gamma$-law treatment in NS mergers has been shown \cite{Bauswein:2010dn,Raithel2021RealisticSimulations}, especially when hyperons become abundant in the  hot and dense matter created \cite{Blacker:2023opp}. Additionally, constraints that are also sensitive to the finite-temperature EoS can be obtained from cooling of NSs \cite{Prakash:1992zng,Page:2004fy,Yakovlev:2004iq, Raduta:2017wpp, Negreiros:2018cho,Fortin:2021umb,Malik:2022jqc,Fortin:2021umb}. Therefore, it is of high interest to extend the EoSs including hyperonic degrees of freedom to the appropriate finite temperatures found in NS mergers and proto-neutron star (PNS) evolution.  

Some progress has been achieved over the last decade, with the development of some hyperonic EoS models that satisfy the astrophysical constraints for the mass, radius and tidal deformability of a NS as well as known experimental low-density constraints (see reviews of \cite{Oertel:2016bki, Burgio:2021vgk,Typel:2022lcx}, the recent works of \cite{Raduta:2021coc,Raduta:2022elz} and references therein).  
However, the uncertainties in the hyperonic effective interactions at high densities and isospin asymmetries are still large and not properly covered by the existing models. Therefore, it is of upmost importance to develop EoSs of homogeneous matter that span a wide range of temperatures ($T = 0 - 100$ MeV), charge fractions ($Y_Q = 0.01 - 0.6$) and baryon densities ($\rho_B = \rho_0/2 - 6\rho_0$), as expected to be achieved in NS mergers or supernovae \cite{Oertel:2016bki}

In Ref. \cite{Kochankovski2022EquationMatter} we developed the hyperonic FSU2H$^{*}$ model, as an extension of the FSU2H one \cite{Tolos:2016hhl,Tolos:2017lgv}, which satisfies the aforementioned mass-radius constraints and at the same time is in agreement with the saturation properties of nuclear matter and finite nuclei, as well as with the constraints on the high-density nuclear pressure coming from heavy-ion collisions \cite{Boguta1977RelativisticSurface,JIANG2007184,LATTIMER_2007,PhysRevLett.86.5647}. The model was then used to study $\beta$-equilibrated matter at finite temperature for neutrino free and neutrino trapped matter.

In the present work we go beyond this previous analysis. On the one hand, we impose weak equilibrium only among hadrons, without taking into account leptons. This is due to the fact that simulations for NS mergers and PNS evolution are performed with cells at a given $T$, $Y_Q$ and $\rho_B$, while implementing the leptons and the transport equations  explicitly. On the other hand, we explore how the uncertainties in the hyperon-nucleon and hyperon-hyperon interactions may influence the properties of the EoS. The hyperon couplings are obtained by fitting the hyperonic potentials to the experimental data, which at present are scarce and subject to big uncertainties. Hence, it is important to determine how these uncertainties propagate to the EoS and, relatedly, their influence on the global properties of the star, such as the composition, masses, radii, tidal deformabilities and moments of inertia. In order to do so, we build two additional parametrizations for the EoS, named hereafter FSU2H$^*$L and FSU2H$^*$U, which cover, respectively, the L(ower) and U(pper) limits of the hyperon potential uncertainties, hence determining a somewhat softer and stiffer hyperonic EoS than the nominal FSU2H$^*$one.

The paper is organized as follows. First, in Section \ref{sec2} we briefly explain the theoretical framework  to construct the EoS at a given $T$, $\rho_B$ and $Y_Q$. Then, in Section \ref{sec3}, we focus on the properties of hyperonic matter, paying a special attention to the effect of the hyperonic uncertainties on the composition of matter and the EoS. Keeping these uncertainties in mind, in Section \ref{sec4} the results for the global properties of both NSs and PNSs are presented, whereas a brief summary of our findings is provided in Section~\ref{sec-conclusions}.

\section{Theoretical framework}\label{sec2}

We consider matter made of baryons ($b$ =  $n, p,\Lambda$, $\Sigma$, $\Xi$) at a given $T$, $Y_Q$ and $\rho_B$. Within the covariant density-functional models, the interaction between baryons is modeled through the exchange of different mesons \cite{Walecka:1974qa}. Our model takes into account the exchange of the scalar mesons $\sigma$ and $\sigma^{*}$, as well as the vector mesons $\omega$, $\rho$ and $\phi$. While the $\sigma$, $\omega$ and $\rho$ mesons mediate the interaction between any
type of baryons in the octet, the $\sigma^{*}$ and $\phi$ mesons mediate the interaction only between particles with non-zero strangeness. The Lagrangian that describes that system can be written in the following form: 
\begin{eqnarray}
{\cal L} &=& \sum_b {\cal L}_b + {\cal L}_m, \nonumber \\
{\cal L}_b &=& \bar{\Psi}_b(i\gamma_{\mu}\partial^{\mu} -q_b{\gamma}_{\mu} A^{\mu} - m_b  
+ g_{\sigma b}\sigma + g_{\sigma^{*}b}  \sigma^{*} \nonumber \\
&-& g_{\omega b}\gamma_{\mu} \omega^{\mu}\nonumber 
- g_{\phi b}\gamma_{\mu} \phi^{\mu} - g_{\rho,b}\gamma_{\mu}\vec{I}_{b} \cdot \vec{\rho\,}^{\mu})\Psi_b, \nonumber \\
{\cal L}_m &=& \frac{1}{2}\partial_{\mu}\sigma \partial^{\mu}\sigma - \frac{1}{2}m^2_{\sigma}\sigma^2 - \frac{\kappa}{3!}(g_{\sigma N}\sigma)^3 - \frac{\lambda}{4!}(g_{\sigma N}\sigma)^4 
 \nonumber \\ 
&+& \frac{1}{2}\partial_{\mu}\sigma^{*} \partial^{\mu}\sigma^{*}  
-\frac{1}{2}m^2_{\sigma^{*}}{\sigma^{*}}^2 \nonumber \\
&-&\frac{1}{4}\Omega^{\mu \nu}\Omega_{\mu \nu}  
+\frac{1}{2}m^2_{\omega} \omega_{\mu} {\omega}^{\mu} +  \frac{\zeta}{4!} g_{\omega N}^4 (\omega_{\mu}\omega^{\mu})^2 \nonumber \\
&-&\frac{1}{4}\vec{R}^{\mu \nu}\vec{R}_{\mu \nu} + \frac{1}{2}m^2_{\rho}\vec{\rho}_{\mu} \cdot \vec{\rho\,}^{\mu}+ 
\Lambda_{\omega}g^2_{\rho N}\vec{\rho_{\mu}}\vec{\rho\,}^{\mu} g^2_{\omega N} \omega_{\mu} \omega^{\mu} \nonumber \\
&-& \frac{1}{4}P^{\mu \nu}P_{\mu \nu}
+\frac{1}{2}m^2_{\phi}\phi_{\mu}\phi^{\mu}-\frac{1}{4}F^{\mu \nu}F_{\mu \nu} \ ,
\label{eq:lagrangian}
\end{eqnarray}
which is split into baryonic (${\cal L}_b$) and mesonic (${\cal L}_m$) contributions. The quantity $m_i$ indicates the mass of particle $i$, $\Psi_b$ is the baryon Dirac field, whereas $\Omega_{\mu \nu} = \partial_{\mu} \omega_{\nu} -\partial_{\nu} \omega_{\mu} $, $\vec{R}_{\mu \nu} = \partial_{\mu} \vec{\rho_{\nu}} - \partial_{\nu} \vec{\rho_{\mu}} $, $P_{\mu \nu} = \partial_{\mu} \phi_{\nu} -\partial_{\nu} \phi_{\mu} $  are the mesonic strength tensors, and $F_{\mu \nu} = \partial_{\mu} A_{\nu} -\partial_{\nu} A_{\mu}$ stands for the electromagnetic one. With $\vec{I}_b$ we represent the isospin operator, while $\gamma^{\mu}$ are the Dirac matrices and $g_{mb}$ labels the coupling of baryon $b$ to meson $m$. We note that, although the contributions of the electromagnetic potential are included in $\cal L$, they do not play a role in the present work, as we are considering charge-neutral objects in the absence of magnetic fields.

As a consequence of the timescale of the weak interaction and the fact that the baryons are in thermal equilibrium in matter, a weak interaction equilibrium can be assumed:
\begin{eqnarray}
&&\mu_{b^0} = \mu_n , \nonumber \\
&&\mu_{b^{-}} = 2\mu_n - \mu_p , \nonumber \\
&&\mu_{b^{+}} = \mu_p,
\label{eq:chemical_potentials_relations}
\end{eqnarray}
where $\mu_b$ labels the chemical potential of the baryon species $b$, and $b^0$, $b^{-}$, $b^{+}$ indicate neutral, negatively charged and positively charged baryons, respectively. As one can see from Eq.~(\ref{eq:chemical_potentials_relations}), only two chemical potentials are independent, $\mu_{p}$ and $\mu_{n}$, which is a consequence of the fact that the strangeness changing weak  reactions are in equilibrium and, hence, the strangeness chemical potential $\mu_{S}$ is zero. Note that we have employed only baryonic chemical potentials as our model only requires to worry about the baryonic EoS. The leptonic part is added explicitly in simulations of NS mergers and supernovae, as mentioned in the Introduction.

In order to obtain the composition and all thermodynamic properties, one needs to solve the Euler-Lagrangian equations of motion. Within the relativistic mean-field (RMF) approximation , the meson field operators are replaced by their expectation values. By labeling  $\bar \sigma = <\sigma>$, $\bar \rho = <\rho_3^0>$, $\bar \omega = <\omega^0>$,  $\bar \phi = <\phi^0>$ and $\bar \sigma^{*} = <\sigma^{*}>$, the following mesonic equations of motion can be obtained: 
\begin{eqnarray}
&&m_{\sigma}^2\bar \sigma + \frac{\kappa}{2}g_{\sigma N}^3 \bar \sigma^2 + \frac{\lambda}{3!}g_{\sigma N}^4 \bar \sigma^3 = \sum_b g_{\sigma b} \rho_b^s , \nonumber \\
&&m_{\sigma^{*}}^2 \bar \sigma^{*} = \sum_{b} g_{\sigma^* b} \rho_{b}^s, \nonumber \\
&&m_{\omega}^2 \bar \omega + \frac{\zeta}{3!}g_{\omega N}^4 \bar \omega^3 + 
2 \Lambda_{\omega} g^2_{\rho N}g^2_{\omega N} \bar \omega \bar \rho^2 = \sum_b  g_{\omega b} \rho_b , \nonumber \\
&&m_{\rho}^2 \bar \rho +  
2 \Lambda_{\omega} g^2_{\rho N}g^2_{\omega N} \bar \omega^{2} \bar \rho = \sum_b \ g_{\rho b}I_{3b} \rho_b , \nonumber \\
&&m_{\phi}^2 \bar \phi = \sum_b g_{\phi b} \rho_b ,
\label{eq:mesons}
\end{eqnarray}
and
\begin{eqnarray}
    &&(i\gamma_{\mu}\partial^{\mu} - q_{q}\,\gamma_{\mu} \,A^{\mu} - m_b^{*} \nonumber \\
    &&- g_{\omega b}\gamma_0 \omega^0 - g_{\phi b}\gamma_{0}\phi^0
    - g_{\rho b} I_{3b}\gamma_0\rho_3^0)\Psi_b = 0,
\label{eq:baryon_motion}    
\end{eqnarray}
for the baryonic one. We use
\begin{equation}
    m_b^{*} = m_b - g_{\sigma b} \sigma- g_{\sigma^{*} b} \sigma^{*},
    \label{eq:meff}
\end{equation}
to label the effective mass of each particle. Note that the scalar $\rho_b^s$ and vector $\rho_b$ densities also appear in the equations of motion. At finite temperature, those are defined as
\begin{eqnarray}
\label{eq:bary_dens}
&&\rho_b = <\bar{\Psi}_b \gamma^0 \Psi_b> \nonumber \\ 
&&=\frac{\gamma_b}{2 \pi^2}\int_0^{\infty}\!\! dk \, k^2 \, (f_{b}(k,T)-f_{\bar{b}}(k,T)) , \nonumber \\
&&\rho_b^s = <\bar{\Psi}_b\Psi_b> \nonumber \\ 
&&= \frac{\gamma_b}{2 \pi^2}\int_0^{\infty} \!\! dk \, k^2 
\, \frac{m^*_b}{\sqrt{k^2+m_b^{*2}}}\left( f_{b}(k,T)+f_{\bar{b}}(k,T) \right), \ \ \ \ 
\end{eqnarray}
where $\gamma_b=2$ labels the spin degeneracy of baryons and 
\begin{equation}
f_{b(\bar{b})}(k,T) =\left[1+\text{exp}\left(\frac{\sqrt{k^2+m_{b(\bar b)}^{*2}}- \mu_{b (\bar b)}^{*}}{T}\right)\right]^{-1} 
\label{eq:distribution}
\end{equation}
is the Fermi-Dirac distribution for a baryon (antibaryon) with an effective mass $m_{b}^{*}=m_{\bar b}^*$, and effective chemical potential:
\begin{equation}
\mu_{b}^{*} = \mu_{b} - g_{{b}\omega}\bar \omega  - g_{{b}\rho} I_{3{b}} \bar \rho- g_{{b}\phi}\bar \phi,
\label{eq:mueff}
\end{equation}
with $\mu_{\bar b}^{*}=-\mu_{b}^{*}$.
The equations of motion of Eqs.~(\ref{eq:mesons}), (\ref{eq:baryon_motion}) and the baryonic densities from Eq.~(\ref{eq:bary_dens}), together with the relations between the different baryon chemical potentials due to weak equilibrium in Eq.~(\ref{eq:chemical_potentials_relations}), are coupled to the baryon conservation number and charge conservation laws given by
\begin{eqnarray}
&&\rho_B = \sum_b \rho_b, \nonumber \\\
&&Y_Q \cdot \rho_B = \sum_{i}q_{i}\rho_i ,
\end{eqnarray}
defining a set of equations that fully determines the composition of matter. With $q_i$ we label the charge of each particle. Once the composition is obtained, it is straightforward to obtain all thermodynamic quantities of interest, such as pressure, energy density, entropy and free energy density,  from the energy-momentum tensor (for details see \cite{Kochankovski2022EquationMatter}).

\subsection{Uncertainties in the hyperonic sector: FSU2H*L and FSU2H*U models}

In the framework of the RMF models, the coupling constants of nucleons to mesons are fitted to reproduce finite nuclei properties and/or bulk nuclear matter properties. When hyperons are also considered in the core of the star, the hyperonic coupling constants are fixed using symmetry principles and constraints derived from hypernuclear data.

The choice of the nucleonic parameters in the FSU2H model was discussed at length in Ref. \cite{Tolos:2016hhl,Tolos:2017lgv}. Given that the present work employs the same parameters for the nucleonic part of the model, we do not repeat the description of these parameters here and just display the values of the coupling constants for nucleons and the meson masses in  Table \ref{tab_2_1}.

While the nucleonic sector is constrained with good accuracy, this cannot be claimed for the hyperonic one, owing to the limited amount of hypernuclear data. For this reason, the hyperon couplings to the vector mesons $\omega$, $\rho$ and $\phi$ are obtained from flavor SU(3) symmetry, considering also the vector dominance model and ideal mixing for the physical $\omega$ and $\phi$ mesons, while 
we leave as free parameters the couplings of the hyperons to the $\sigma$ meson ($g_{\sigma \Lambda}, g_{\sigma \Xi}, g_{\sigma \Sigma}$), and the coupling of the $\Lambda$ hyperon to the $\sigma^{*}$ meson ($g_{\sigma^{*} \Lambda}$). 
The hyperon-$\sigma$ coupling constants are determined by reproducing the potential felt by the hyperon in symmetric nuclear matter, whereas the $\Lambda-\sigma^*$ one results from obtaining the $\Lambda \Lambda$ bond energy extracted from double $\Lambda$ hypernuclei. We recall that the potential felt by a hyperon $i$ in $j$-particle matter is given by:
\begin{equation}
%\begin{eqnarray}
%&& 
U_i^{(j)}(\rho_j) = 
%\nonumber \\
%&& 
-g_{\sigma i}\bar{\sigma}^j -g_{\sigma i}\bar{\sigma^{*}}^j + g_{\omega i}\bar{\omega}^j+g_{\omega i}\bar{\rho}^j I_{3i} +g_{\phi i}\bar{\phi}^{(j)}. 
%\end{eqnarray}
\end{equation}

\begin{table*}
\caption{Parameters of the model FSU2H$^*$ for the nucleon coupling constants and mesons masses. The mass of the nucleon is equal to $m_N = 939$ MeV.}
{\small
\label{tab_2_1}       % Give a unique label
\begin{tabular}{cccccccccccccc}
\hline\noalign{\smallskip}
$m_{\sigma}$  & $m_{\omega}$ & $m_{\rho}$&$m_{\sigma^{*}}$ & $m_{\phi}$ &$g_{\sigma N}^2$  & $g_{\omega N}^2$ & $g_{\rho N}^2$ & $\kappa$ & $\lambda$ & $\zeta$ & $\Lambda_{\omega}$ \\
(MeV) &  (MeV)&  (MeV)&  (MeV)& (MeV) & & & & (MeV) & & &\\
\noalign{\smallskip}\hline\noalign{\smallskip}
497.479 & 782.5 & 763 & 980  & 1020 & 102.72 & 169.53 & 197.27 & 4.0014 & -0.0133 & 0.008 & 0.045 \\
\noalign{\smallskip}\hline
\end{tabular}
}
\end{table*}

\begin{table*}
\centering
\caption{The ratios of the hyperon couplings to $\sigma$, $\omega$ and $\rho$ mesons with respect to the nucleon ones, as well as the ratios of the couplings of hyperons to $\sigma^*$ and $\phi$ with respect to the $\sigma$-nucleon and $\omega $-nucleon ones, respectively, due to the OZI rule. Note that the couplings to the $\sigma$ and $\sigma^{*}$ mesons depend on the model, so we list the original values of the FSU2H$^*$ model, together with those of the FSHU2$^*$U and FSHU2$^*$L models, the latter ones labeled with (U) and (L), respectively. The other ratios are the same for the different models.}

\label{tab_2_2}       % Give a unique label
\begin{tabular}{ccccccccccc}
\hline\noalign{\smallskip}
$Y$ & $R_{\sigma Y}$ & $R_{\sigma Y} (U)$ & $R_{\sigma Y} (L)$ & $R_{\sigma^{*}Y}$ & $R_{\sigma^{*}Y} (U)$ & $R_{\sigma^{*}Y} (L)$ &$R_{\omega Y}$ & $R_{\rho Y}$ & $R_{\phi Y}$ \\
\noalign{\smallskip}\hline\noalign{\smallskip}
$\Lambda$ & $0.6113$ & $0.6048$ & $0.6178$  & $0.2812$ & $0.2309$ & $0.4954$ & $2/3$ & $0$ & $-\sqrt{2}/3$  \\
$\Sigma$ & $0.4673$  & $0.4085$  & $0.5132$  & $0.2812$ & $0.2309$ & $0.4954$ & $2/3$ & $1$ & $-\sqrt{2}/3$  \\
$\Xi$ & $0.3305$ & $0.2938$ & $0.3325$  & $0.5624$ & $0.4618$ & $0.9908$ & $1/3$ & $1$ & $-2\sqrt{2}/3$  \\
\hline\noalign{\smallskip}
\end{tabular}
\label{table:hypparam}
\end{table*}

In our previous work \cite{Kochankovski2022EquationMatter} the free parameters of the FSU2H$^*$ model were determined employing the central value of the most accepted determinations of the hypernuclear potentials. The $\Lambda$ potential at saturation density is by far the best constrained value. According to \cite{PhysRevC.38.2700}, the value that reproduces the bulk of
$\Lambda$ hypernuclei binding energies  is $U_{\Lambda} = -28$ MeV. The potentials for the other hyperons in symmetric nuclear matter are not so well constrained. In \cite{Kochankovski2022EquationMatter} we considered the latest value obtained in Ref.~\cite{Friedman2021ConstraintsEmulsion} for the $\Xi$ potential of $U_{\Xi} = -24$ MeV, and we adopted $U_{\Sigma} = 30$ MeV \cite{FRIEDMAN_2007} for the $\Sigma$ potential. Moreover, we considered a $\Lambda \Lambda$ bond energy in $\Lambda$ matter at a density $\rho = \frac{\rho_0}{5}$ of $\Delta B_{\Lambda \Lambda}$($\rho_0/5$) = $-0.67$ MeV) \cite{Takahashi:2001nm,Ahn:2013poa} to derive the $\sigma^{*}\Lambda$ coupling $g_{\sigma^{*}\Lambda}$. 

However, in the present paper we want to explore the uncertainties in the hyperonic sector and how they propagate to the EoS and the global properties of compact stars. Therefore, as noted before, we construct two extreme models, FSU2H$^*$U and FSU2H$^*$L, that were obtained by taking the hyperonic potentials to be equal, respectively, to the upper and lower limits of their uncertainty bands. The band is pretty narrow for the $\Lambda$ hyperon, $U_{\Lambda} = (-30, -25)$ MeV, while the other hyperonic potentials have wider uncertainties, that is,  $U_{\Xi} = (-24,-10)$ MeV, $U_{\Sigma} = (10, 50)$ MeV \cite{Gal_2016}, $\Delta B_{\Lambda \Lambda} = (-6, 0)$ MeV \cite{Gal_2016,Guo:2021vsx,Friedman2021ConstraintsEmulsion}. A more detailed justification for our choice can be found in Appendix \ref{sec:app}.  In Table \ref{tab_2_2} we show the values of the hyperonic constants to the mesons for the three models that we will use hereafter: FSU2H$^*$, FSU2H$^*$U and FSU2H$^*$L. In the case of the $\sigma$, $\omega$ and $\rho$ mesons, those hyperon couplings are given as a ratio to the corresponding couplings of the nucleon, namely $R_{i Y}=g_{iY}/g_{iN}$ with $i={(\sigma,\omega,\rho)}$, whereas for the $\sigma^*$ and the $\phi$ mesons we give $R_{\sigma^* Y}=g_{\sigma^* Y}/g_{\sigma N}$ and $R_{\phi Y}=g_{\phi Y}/g_{\omega N}$, since their nucleonic couplings are zero due to the OZI rule.

\section{Composition and the Equation of State at finite \textit{T}}
\label{sec3}

In the present section we aim at exploring the effect of the hyperonic uncertainties in the composition and the EoS of matter for the wide range of $T$, $Y_Q$, and $\rho_B$ described in the Introduction, using the two newly constructed parametrizations. Thus, we consider matter at three different temperatures: low ($T = 1$ MeV), moderate  ($T = 20$ MeV) and high temperature ($T = 80$ MeV) and three different charge fractions: $Y_Q = 0.01$, $Y_Q = 0.25$ and $Y_Q = 0.5$, and examine the composition and the corresponding EoSs as functions of $\rho_B$. However, we note that the composition patterns are sensitive to the interplay between the different hyperonic potentials and, consequentily, the uncertainty region of each individual hyperonic species is only partially covered by the FSU2H$^*$U and FSU2H$^*$L models (see Appendix \ref{sec:app} for more details).

\subsection{Composition}\label{subsec2-2}

\begin{figure*}
\centering
\includegraphics[width=1.0\textwidth]{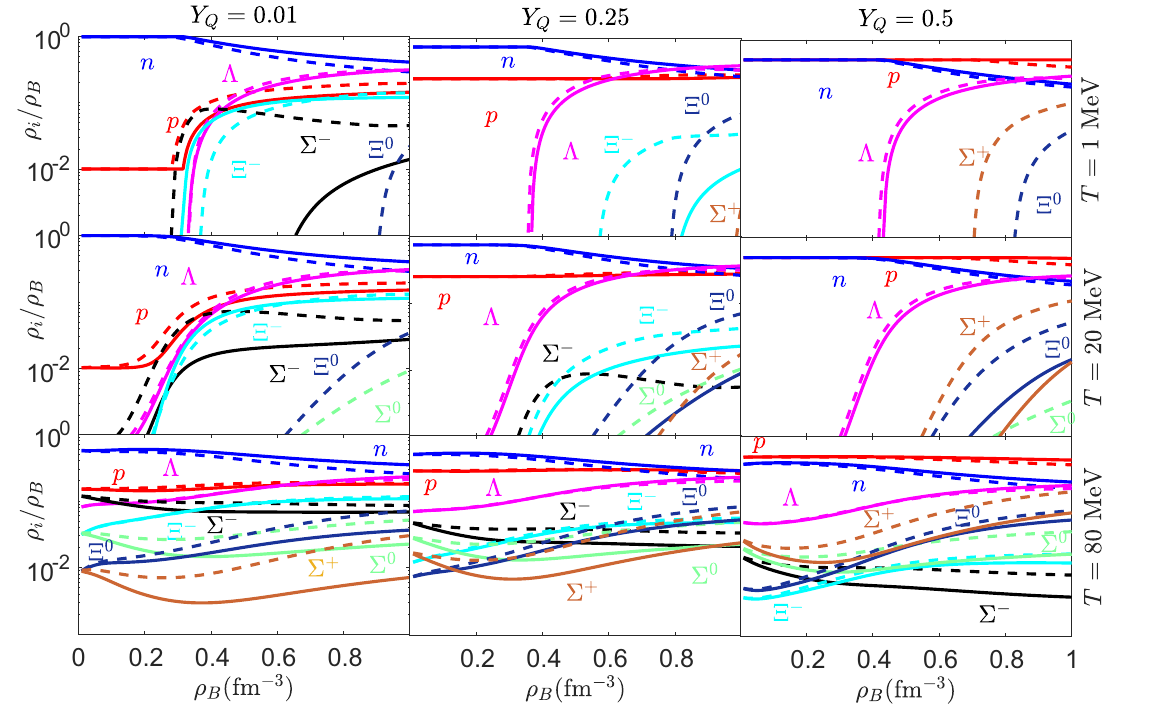}
\caption{Composition of baryonic matter for different charged fractions (columns) and temperatures (rows). Solid lines correspond to calculations with the FSU2H$^*$U model, while dashed lines correspond to those with the FSU2H$^*$L one.}
\label{fig:1}
\end{figure*}

In Fig. \ref{fig:1} we show the different baryonic fractions as functions of the baryon density for $T=1$ MeV (upper panels), $T=20$ MeV (middle panels) and $T=80$ MeV (lower panels), and for constant charge fractions $Y_Q=0.01$ (left column), $Y_Q=0.25$ (middle column) and $Y_Q=0.5$ (right column).
We note that, when nucleons are the only baryons present, the composition is trivially determined by the charge fraction, as for any temperature the partial densities of the neutrons and protons are $\rho_n = (1-Y_Q)\rho_B$ and $\rho_p = Y_Q\rho_B$ (hence $x_n = (1-Y_Q)$ and $x_p = Y_Q$). This is no longer the case when hyperons are included, and the abundance of the different particles is a function not only of the baryon density but also of the temperature.
From the figure, one can make the following observations. First, at low temperature, the baryon density at which a hyperon species appears strongly depends on the hyperonic potentials. This is especially important for the $\Sigma$ hyperons since the uncertainty in their potential is the largest. For the charged fraction $Y_Q = 0.01$, due to the big difference in the chemical potential of the neutron and the proton, the $\Sigma^{-}$ hyperon can appear even before $\Lambda$ within the FSU2H$^*$L model. On the contrary, for the FSU2H$^*$U model the $\Sigma^{-}$ hyperons start appearing at densities greater than $0.6$ fm$^{-3}$ due to the highly repulsive potential. 
Also, at small charge fractions, the increased abundance of negative hyperons with density induces the fraction of protons to increase as well. However, this process can happen only at small charged fractions. If one analyses the composition of matter at low temperatures ($T=1$ MeV) and larger charge fractions ($Y_Q = 0.25$ and $Y_Q = 0.5$), it can be noticed that the proton abundance does not change significantly and the appearance of negative hyperons is hindered. 

Neutrons, on the other hand, can always be replaced with non-degenerate low energetic $\Lambda$ at sufficiently large densities, so in all of the $Y_Q$ cases, their abundance is significantly reduced. However, the density at which this reduction begins depends on $Y_Q$. When the matter is highly isospin asymmetric, this process starts before $\rho_B = 0.3$ fm$^{-3}$, while when matter is isospin symmetric, it starts at densities greater than $\rho_B = 0.4$ fm$^{-3}$.

The previous analysis at low temperature still holds for moderate temperatures ($T = 20$ MeV). However, there are two main differences. First, due to thermal effects, the hyperon fractions $x_Y = \rho_Y/\rho_B$, where $Y$ represents any hyperon, start to be significant ($x_Y>10^{-3}$) at lower baryon density, and the evolution of their abundances with density is smoother than at low temperatures. Second, for a given density, more hyperonic species can be present, especially when the hyperon potentials are lower. Still, at low temperatures, the $\Lambda$ hyperon contribution does not change significantly with the EoS employed. The reason lies in the fact that the $\Lambda$ potential is better constrained than that of the other hyperonic species.

The composition of baryonic matter significantly changes at high temperatures ($T=80$ MeV). It is noticeable that all hyperons have an appreciable abundance at any density. Also, it is interesting to observe that, in all $Y_Q$ cases, the neutron abundance is no longer decreasing monotonically with increasing density as for lower temperatures. The difference in the abundance of the $\Sigma$ hyperons obtained with the FSU2H$^*$U or FSU2H$^*$L models observed at lower temperatures is especially manifest for the $\Sigma^{+}$ at this high temperature of $T=80$ MeV. Being the only positively charged hyperon, its appearance is key for the reduction of the proton contribution. This effect is clearly seen in the $Y_Q = 0.5$ case, where the $\Sigma^{+}$ abundance in the FSU2H$^*$L model reaches that of the $n$ and the $\Lambda$ at high densities. 

\begin{figure*}
\centering
\includegraphics[width=1.0\textwidth]{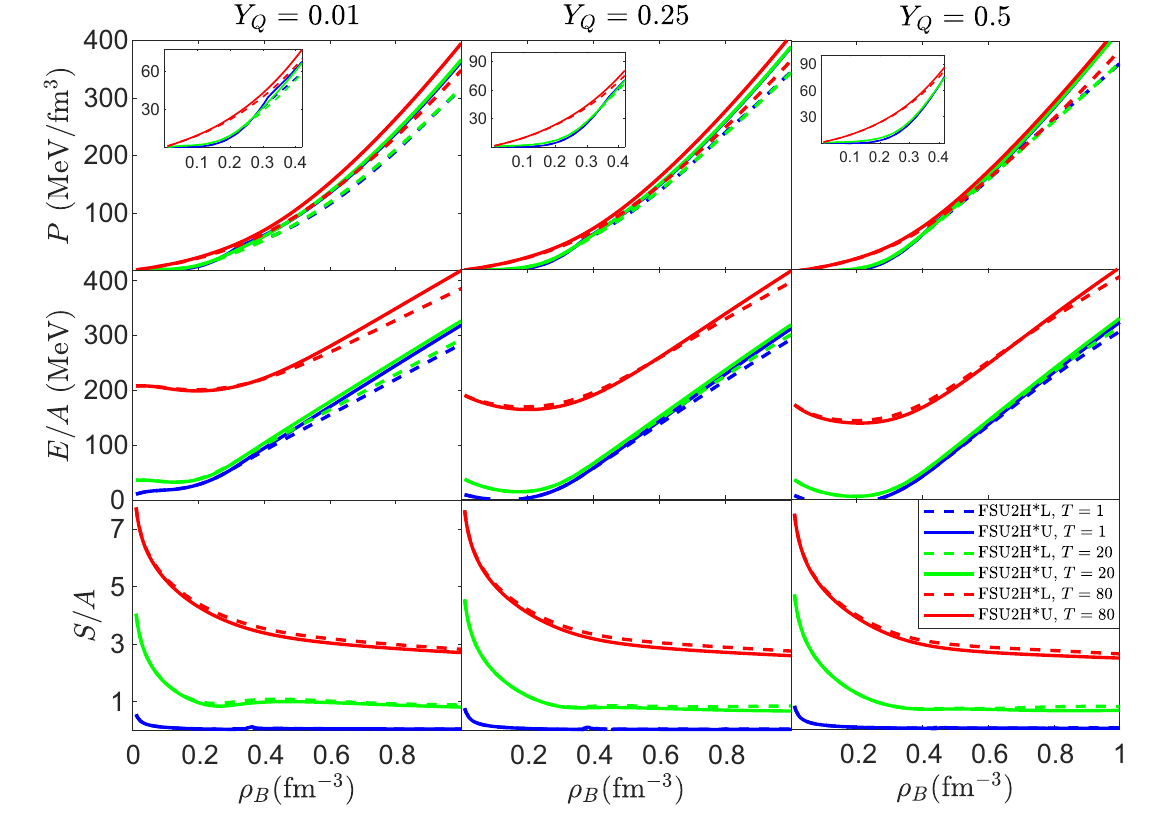}
    \caption{EoS of hypernuclear matter for different charged fractions and temperatures. Solid lines correspond to calculations with FSU2H$^*$U model, while dashed lines correspond to ones with FSU2H$^*$L. Blue lines are calculations done at $T=1$ MeV, green lines are performed at $T=20$ MeV, while red ones are at $T = 80$ MeV. The pressure $P$ at low densities $\rho_B$ up to $0.4 {\rm fm^{-3}}$ is shown with inset plots in the upper panels.}
    \label{fig:2}
\end{figure*}

\subsection{Equation of State}\label{subsec2-3}

In this subsection we present the EoS for the two models, FSU2H$^*$U and FSU2H$^*$L. In particular, we focus on the behavior of the pressure, $P$, energy per particle, $E/A$ and entropy per particle, $S/A$, as functions of the baryon density for the same $T$ and $Y_Q$ cases discussed above. All plots are combined in Fig. \ref{fig:2}, where the different colors stand for different temperatures and the solid (dashed) lines refer to the results of the FSU2H$^*$U (FSU2H$^*$L) model.

It is clear that the effects of the hyperonic uncertainties are more visible in the pressure. At high density, this effect can even be more important than the temperature corrections to the pressure. To illustrate this point, we focus first on the $P(\rho_B)$ dependence at $Y_Q = 0.01$. For densities larger than $\rho_B = 0.6\ \ \text{fm}^{-3}$, the $T=80$ MeV FSU2H$^*$L pressure (dashed red line) is lower than the $T=1$ MeV FSU2H$^*$U one (solid blue line). This is more evident as we move to more isospin symmetric matter, that is, to larger values of $Y_Q$. As matter becomes proton richer, the abundance of hyperons decreases. This means that high density matter is more degenerate, leading to pressure-density curves for different temperatures that do not differ noticeably in the case of the stiffest model FSU2H$^*$U. In contrast, as mentioned when describing the composition, the softest FSU2H$^*$L model allows for a significant abundance of positively charged $\Sigma^{+}$ hyperons, which replace highly energetic protons. This reduction of the matter degeneracy makes the FSU2H$^*$L pressure-density curves at different temperatures to be distinguishable.

As for the density dependence of the energy per particle, $E/A$, it is clear from the middle panels of Fig.~\ref{fig:2} that the differences between the two models are larger for high densities and high temperatures. When the charge fraction is low ($Y_Q = 0.01$), more hyperons can appear, so the difference between the FSU2H$^*$L and FSU2H$^*$U  models becomes more noticeable.

Finally, in the lower panels of Fig.~\ref{fig:2} we show the entropy per particle, $S/A$, as a function of the baryon density. We observe that $S/A$ does not show a clear visible dependence on the model. This means that the uncertainty of the hyperon couplings will not significantly influence the temperature profile of the early 
evolution of the star, as we will explicitly show in Section \ref{sec4}. However, we remind that the temperature profiles computed for stars at constant $S/A$ change drastically, acquiring a plateau-like behavior, as soon as hyperons appear in matter, as discussed in \cite{Raduta:2020fdn,Kochankovski2022EquationMatter}. These are not, however, contradictory statements. When it is energetically possible, the most degenerate nucleons are converted into hyperons and, therefore, a new Fermi sea is being filled. Thus, the presence of hyperons influences strongly the entropy, which acquires sensibly larger values than in their absence, and develops the plateau-like structure as a function of $\rho_B$ seen in Fig.~~\ref{fig:2}. However, once hyperons are present, the effect of the hyperonic potential uncertainties in $S/A$ turns out to be quite mild.

\section{Influence of the hyperonic uncertainties in astrophysical observables}
\label{sec4}

In this section we aim at showing how the hyperonic uncertainties propagate from the composition and EoS of NS matter to  astrophysical observables, such as the mass, radius, tidal deformability and moment of inertia.

The masses ($M$) and radii ($R$) of PNSs and cold NSs are obtained from solving the TOV equations assuming $\beta^{-}$ stable conditions for baryons and leptons (electrons and muons), with trapped or untrapped neutrinos, respectively. In particular, the maximum mass that the model predicts is especially sensitive to the appearance of the hyperons, as the EoS becomes softer.   
 
In recent years, due to the newest gravitational wave detectors \cite{lasky_2015} and the planned ones \cite{Bailes:2021tot}, the tidal deformability $\lambda$ of the star has become a particularly important quantity. It measures the induced quadrupole moment in one star in response to the tidal field of the companion and can be obtained from the tidal Love number $k_2$ as 
\begin{equation}
    \lambda = \frac{2k_2R^5}{3G},
\end{equation}
where $G$ is the gravitational constant and $R$ is the radius of the star. The $k_2$ number is obtained from a first-order ordinary differential equation \cite{Hinderer:2007mb,Hinderer_2010}, solved self-consistently with the
integration of the TOV equations. It is useful to define the dimensionless tidal deformability ($\Lambda$) as 
\begin{equation}
    \Lambda = \frac{\lambda}{G^4 M^5}.
\end{equation}
The tidal deformability constraints are usually given for the chirp mass of a binary NS system, which is extracted from the masses of the individual stars $M_1$ and $M_2$ through the relation:
\begin{equation}
   {\cal M} = \frac{(M_1M_2)^{\frac{3}{5}}}{(M_1+M_2)^{\frac{1}{5}}}.
\end{equation}

Another interesting quantity is the moment of inertia of the star $I$, that can be computed by solving the structure of a rotating NS using the Hartle–Thorne approach \cite{1967ApJ...150.1005H, 1968ApJ...153..807H}.
Similar to the tidal deformability, the moment of inertia also puts simultaneous constraints on the mass and the radius of the star. Still, there is no independent measurement of this quantity at present. The first measurements with an accuracy of 10\% can be expected in the next decade \cite{Lattimer_2016}.

Given that our goal is to study the effect of hyperons and their uncertainties, we also calculate the so-called strangeness number of the star. This number is related to the abundance of hyperons in the interior of the star, and it can be computed as
\begin{equation}
 \mathcal{S} = 4 \pi \int_0^R \frac{r^2 \sum_i n^s_{i}}{\sqrt{1-\frac{2 G M(r)}{r}}} dr,    
\end{equation}
where $n^s_{i} = s_i \rho_i$ is the strangeness density for each baryon species $i$, with $s_i$ standing for the corresponding strangeness quantum number. As the strangeness number is large, we will present results for the ratio $s = \mathcal{S}/N_B$, which is known as the normalized strangeness number, where $N_B$ is the total baryon number given by
\begin{equation}
 N_B = 4 \pi \int_0^R \frac{r^2 \sum_i \rho_{i}}{\sqrt{1-\frac{2 G M(r)}{r}}} dr.   
\end{equation}

\subsection{Results at zero temperature}

Since most of the astrophysical constraints that are available are obtained for already evolved cold NSs, we dedicate this subsection to present our $T=0$ results, while the next subsection will be devoted to the discussion of the finite temperature results that are especially important for the relativistic simulations of NS mergers and supernovae.

The masses, radii, tidal deformabilities and moments of inertia for neutrinoless $\beta^{-}$ stable NSs at $T=0$ are computed with our three models, that is, FSU2H$^*$, FSU2H$^*$L and FSU2H$^*$U. Our core EoSs have to be matched with an EoS for the inner crust and an EoS for the outer crust. In Ref.~\cite{10.3389/fspas.2019.00013} the EoS for the inner crust with the FSU2H interaction was computed, allowing us to have a unified EoS description of the core-inner crust of the star. For the outer crust, we adopt the widely used EoS of Baym-Pethick-Sutherland (BPS) \cite{1971ApJ...170..299B}, which is well constrained by nuclear physics data.

\begin{figure}
\centering
\includegraphics[width=0.45\textwidth]{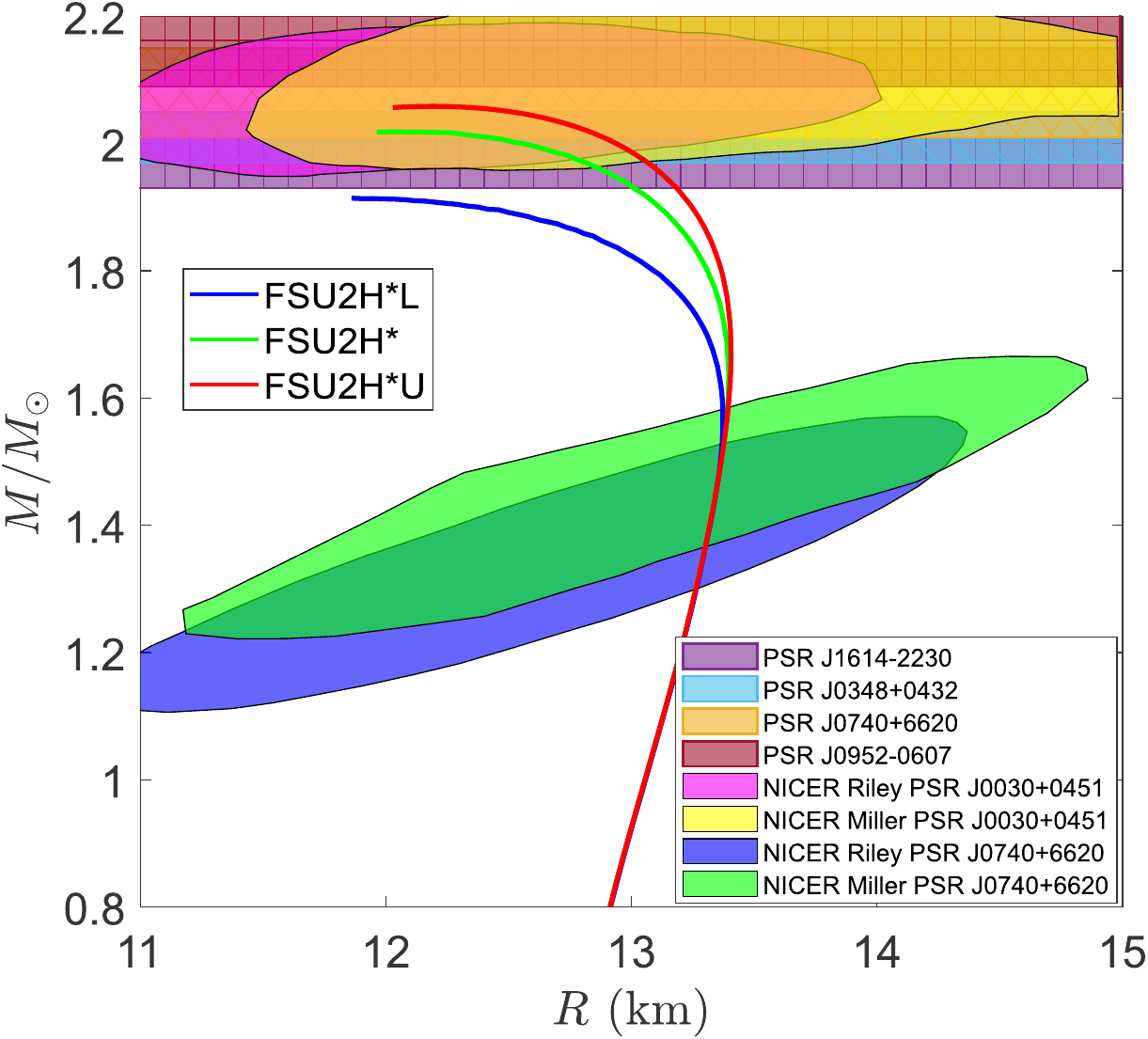}
\caption{$M(R)$ relation obtained with the models FSU2H$^{*}$, FSU2H$^{*}$L and FSU2H$^{*}$U along with all with the observational constraints obtained from 2$ \protect M_{\odot}$ observations \protect \cite{Demorest2010ShapiroStar,Antoniadis:2013pzd,Fonseca2016,Cromartie2020RelativisticPulsar} and NICER measurements of radii within a 1$\sigma$ error band \protect \cite{Riley:2019yda,Miller:2019cac,Riley:2021pdl,Miller:2021qha}.}
\label{fig:3}
\end{figure}

In Fig.~\ref{fig:3} we show the mass-radius $M(R)$ relation for the three models, i.e., FSU2H$^*$ (green line), FSU2H$^*$L (blue line) and FSU2H$^*$U (red line), together with the constraints for masses coming from 2$M_{\odot}$ observations \cite{Demorest2010ShapiroStar,Antoniadis:2013pzd,Fonseca2016,Cromartie2020RelativisticPulsar,Romani:2022jhd} and NICER measurements of radii \cite{Riley:2019yda,Miller:2019cac,Riley:2021pdl,Miller:2021qha}. We note that, as other hyperonic EoSs, the developed EoSs are in tension with the measurement of the heavy pulsar PSR J0952-0607 \cite{Romani:2022jhd}.

We focus on the most massive NSs, with $M>1.5M_{\odot}$, as the effect of hyperons is more pronounced. As previously discussed, the original FSU2H$^*$ can explain all other aforementioned $M$-$R$ constraints. However, the FSU2H$^*$L falls short in describing NSs with $2M_{\odot}$ masses. The reduction of the hyperonic potentials in this model softens the EoS significantly and the maximum mass the model can support is $1.92 {\rm M}_{\odot}$. On the contrary, the FSU2H$^*$U model, due to the more repulsive hyperonic potentials, gives rise to a higher maximum mass of $2.06 {\rm M}_{\odot}$. Although the hyperonic uncertainties have a significant impact on the value of the maximum NS mass, the corresponding radii turn out to be quite similar. For example, the FSU2H$^*$L model predicts its maximum star mass to have a radius of around  ${R = 11.9}$ km, while the FSU2H$^*$U model predicts a radius of around $R = 12.1$ km for its maximum mass configuration. However, the hyperonic uncertainties have a significant impact on the radius for a fixed mass star in the range $1.6$-$1.9 M_{\odot}$. For instance, the difference between the radius of the FSU2H$^*$L maximum star mass and that of a star with the same mass in the FSU2H$^*$U model is larger than $1.2$ km, which corresponds to an uncertainty of around $10\%$.    

\begin{figure}
\centering
\includegraphics[width=0.45\textwidth]{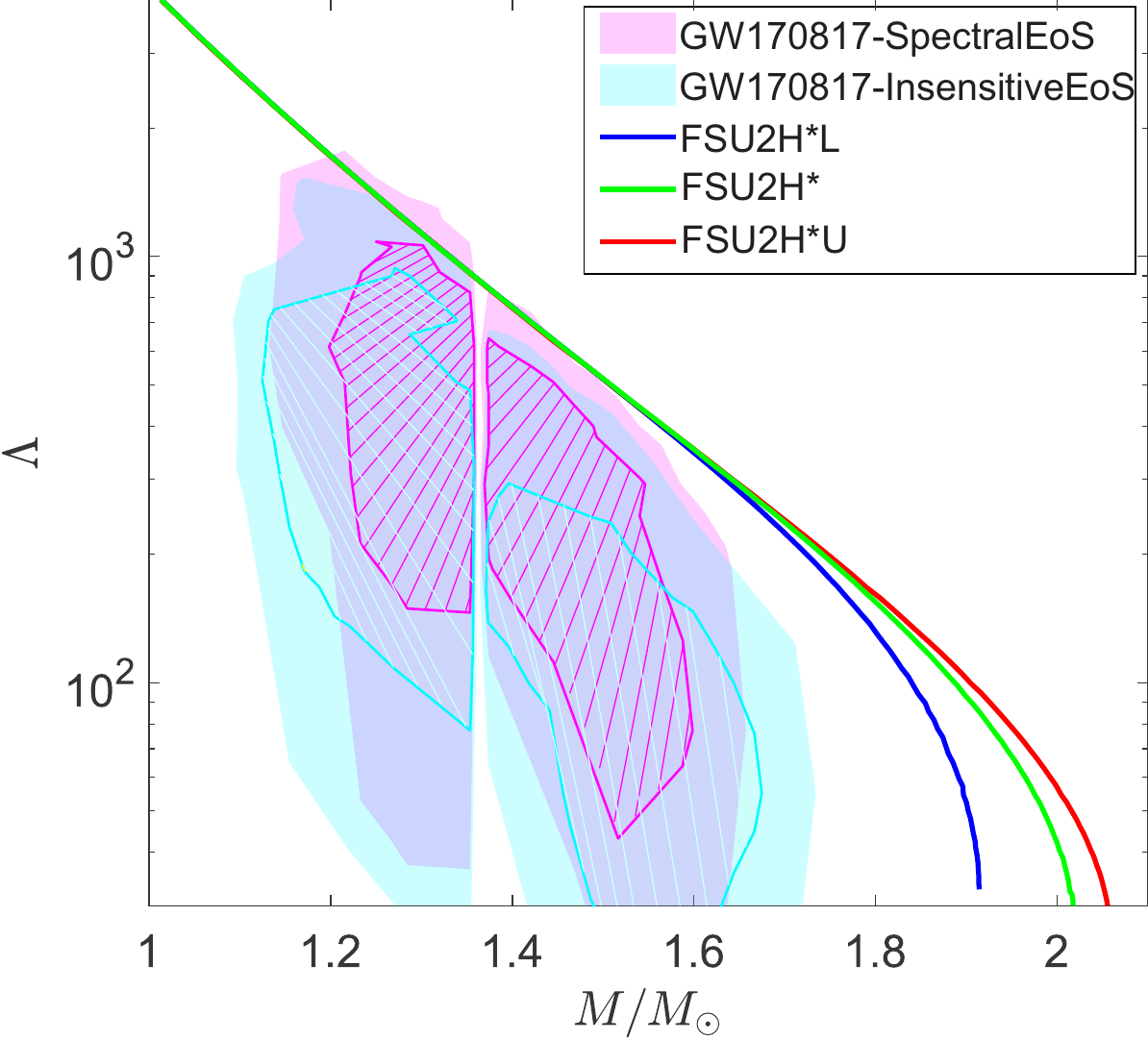}
\caption{$\Lambda(M)$ relation obtained with the FSU2H$^*$, FSU2H$^*$L and FSU2H$^*$U models along with the observational tidal constraint using the universal relations (EoS insensitive) and the spectral EoS obtained from GW170817 event. The filled boundaries represent the values at $95\%$ confidence level, while the hatched ones represent the values at $68\%$ confidence level. For details, see Ref. \protect \cite{MUSES:2023hyz}.}

\label{fig:4}

\end{figure}

In Fig.~\ref{fig:4} we show the dimensionless tidal deformability $\Lambda$ of the star as a function of the mass of the star (note the logarithmic scale) along with constraints presented in \cite{MUSES:2023hyz}. Even if at the upper edge, our EoSs are in agreement with the constraints, which so far cover the region of low-medium sized NS masses, where hyperons are not yet present. The trend of the tidal deformability for most of the region is exponentially decreasing. However, from a certain mass onwards the exponential trend breaks down and the tidal deformability starts decreasing at a faster rate. This corresponds to the mass in the $M$-$R$ diagram from which a small increase produces a significant decrease in the radius, namely around $1.7{\rm M}_{\odot}$ in the FSU2H$^*$L model and $1.9{\rm M}_{\odot}$ in the FSU2H$^*$ and FSU2H$^*$L ones. As a consequence, we find the effect of the hyperon uncertainties on the tidal deformability to be significant for the most massive stars $M \gtrsim 1.8{\rm M}_{\odot}$. More quantitatively, we observe that for a star with a mass $M = 1.92{\rm M}_{\odot}$, the FSU2H$^*$L model gives $\Lambda \approx 34$, while the FSU2H$^*$U predicts $\Lambda \approx 95$. This large discrepancy for the tidal deformability is correlated to the different values of the radii for a given mass \cite{PhysRevD.82.024016}. We thus conclude that hyperonic uncertainties have a significant impact on the tidal deformability for the most massive stars.

\begin{figure}
\centering
\includegraphics[width=0.45\textwidth]{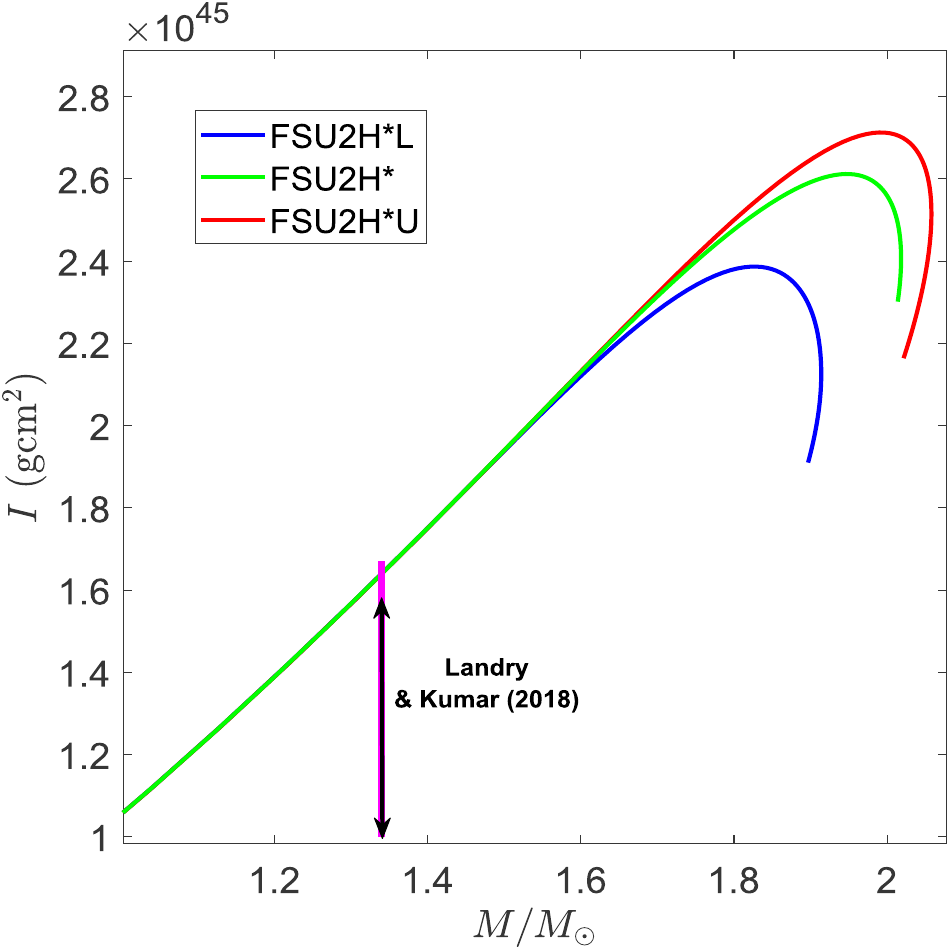}

\caption{$I(M)$ relation obtained with the models FSU2H$^*$, FSU2H$^*$L and FSU2H$^*$U along with the observational constraints obtained from Ref.~ \protect \cite{Landry_2018}.
}

\label{fig:5}

\end{figure}

In Fig.~\ref{fig:5} we show the moment of inertia $I$ as function of the mass of the star computed with our models, together with the two constraints for $I$ given in Ref.~\cite{Landry_2018}, which do not correspond to direct measurements, but are derived from universal relations among NS observables employing the more restrictive and less restrictive range of values for the tidal deformability reported in \cite{LIGOScientific:2018hze} and \cite{LIGOScientific:2017vwq}, respectively. 
We note that our EoSs do not satisfy the tighter constraint. 
Within error bars, our results are almost compatible with the recent model-dependent Bayesian analysis of Ref.~\cite{Miao:2021gmf}.
A direct, independent, measurement of the moment of inertia, which might become available in the next years from radio observations of the double pulsar
PSR J0737-3039, would pose a more trustable constraint on the EoS models (\cite{Bejger_2005, Lattimer_2005,Greif_2020, Hu_2020}). 

We also note that the moment of inertia increases linearly with the mass of the star in the region between $1.0 M_{\odot}$ and $1.6M_{\odot}$. For larger masses this trend is broken, with the stiffer FSU2H$^*$U favouring this linear trend even for masses up to $1.8 M_{\odot}$. Whereas the qualitative trend of the three curves is similar, their quantitative predictions for $I$ for the most massive stable stars can differ by around $10\%$.

\begin{figure}
\centering
\includegraphics[width=0.45\textwidth]{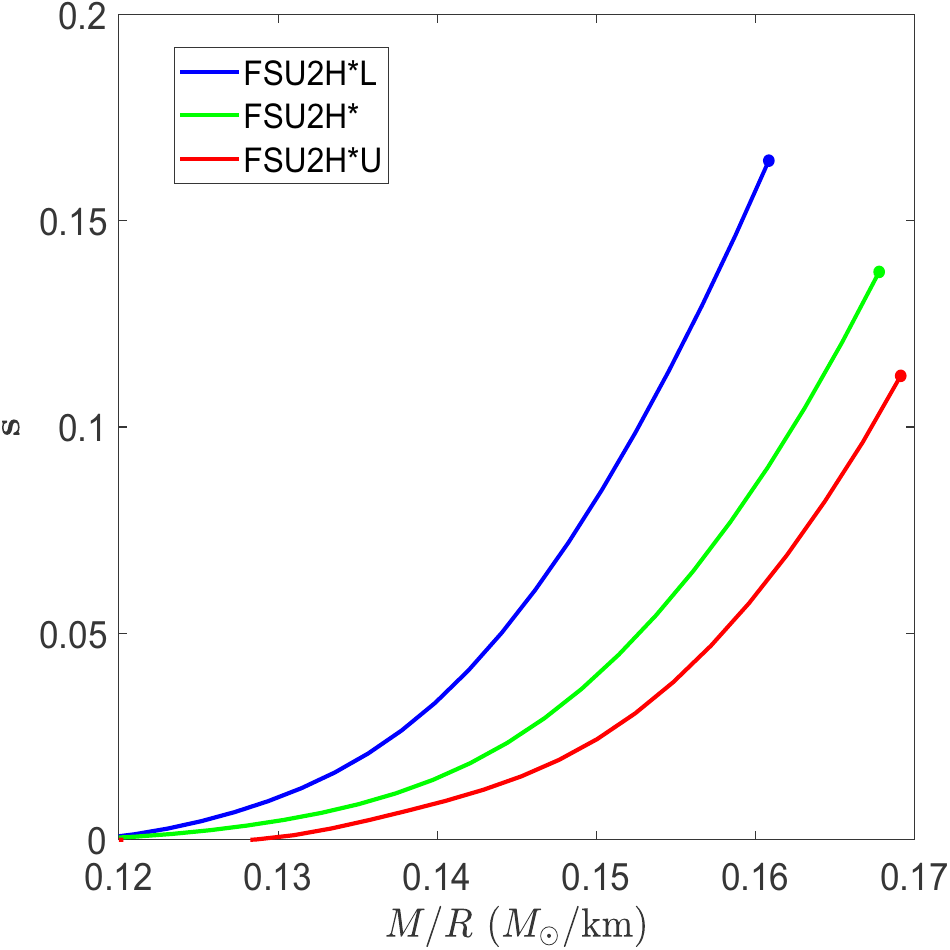}
\caption{Normalized strangeness number $s$  as a function of compactness for the FSU2H$^*$, FSU2H$^*$L and FSU2H$^*$U models. Each line ends with a dot that indicates the maximum mass star configuration that is stable.}
\label{fig:6}
\end{figure}

We finish this subsection showing the normalized strangeness number as a function of the compactness of the star in Fig.~\ref{fig:6}. As expected,  the strangeness number increases monotonically with the compactness for all different parametrizations. This is due to the fact that higher compactness results in both the average density and the central density of the star to be larger and, hence, strange particles can be produced more easily. It is also interesting to note that for, a given value of compactness,  the softer the model is, the larger the strangeness number becomes. This is due to the fact that the central density for soft models is significantly larger with respect to the stiff ones.

\begin{figure*}
\centering
\includegraphics[width= 0.9\textwidth]{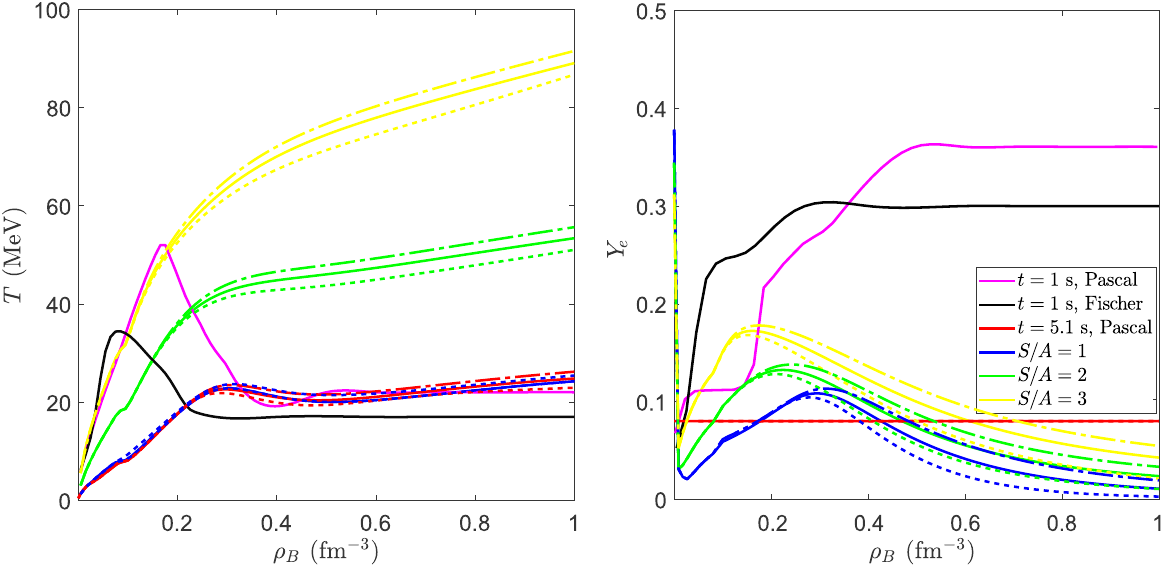}
\caption{Temperature  (left plot) and electron fraction profiles (right plot) of the stars used in our calculations. Solid lines represent the results with FSU2H model, dashed lines represent FSU2H$^*$L ones and dashed-dotted lines are results with the FSU2H$^*$U model. We note that profiles inspired by simulations at $t=1$ s are the same for all three models.}
\label{fig:7}
\end{figure*}

\subsection{Results at finite temperature}

Here we present results for stars in $\beta^{-}$ equilibrium at finite temperature. The focus will be on cases important for the evolution of PNSs. Here, the temperature ($T=T(\rho_B)$) and electron/lepton ($Y_e=Y_e(\rho_B)$) profiles are essential inputs to obtain the properties of the star. For the early evolution of the star, at the so-called deleptonization phase (approximately one second after the supernova explosion), we will adopt a similar approach to the one in Ref. \cite{Marques_2017}, where the temperature profiles are inspired by results of full relativistic simulations that also account for the neutrino transport. However, instead of assuming a constant lepton fraction throughout the star, we will also use lepton profiles inspired by simulations, so as to make the calculations more realistic, in particular, we will take profile inspired by the results in Ref. \cite{Raduta:2021coc} ($t=1$ s, Pascal) and a profile inspired from results obtained with the FSU2H EoS, with the code and microphysics as described in \cite{Fischer_2010,Fischer_2012, Fischer:2016boc, Fischer_2020, Fischer:2020vie} ($t=1$ s, Fischer). We will also perform calculations at fixed entropy per baryon ($S/A = 1$), and at constant lepton fraction ($Y_e = 0.08$) throughout the star ($t=5.1$ s, Pascal). This profile accounts for conditions in the star of around five seconds after the explosion, and is motivated by the study of Ref. \cite{Pascal:2022qeg}, which found that the convective motions in the star enabled the isentropic and isoleptonic profiles to be achieved much faster. Finally,
for the later stage of the evolution of the star (approximately tens of seconds after the explosion), we will focus on cases where the neutrinos have already diffused out from the star and the entropy of the star slowly reduces, as it was recently shown in Ref. \cite{Schneider_2020} that these stars are relevant for black hole formation in a failed core collapse supernovae. In the present work, we will cover this later evolution situation by considering neutrino-free matter at three constant entropy profiles, $S/A=1,2$ and $3$.

Before showing the results, a word of caution is necessary. Similarly to the $T=0$ case, for temperatures lower than $T\approx15$ MeV and densities below $\rho_0/2$ \footnote{the exact transition point depends on both $T$ and $\rho_B$ as well as the model that is used (see Ref.~\cite{Hempel_2010})} cluster structures are present in the star. The theoretical framework suitable under these conditions is the extended Nuclear Statistical Equilibrium (NSE) \cite{Furusawa:2011wh, Furusawa:2013rta, Furusawa:2016tdj, Furusawa:2017auz, Shen:2010pu, Shen:2010jd, Shen:2011kr, Shen:2011fc, Hempel_2010, Typel:2009sy, Gulminelli:2015csa, Raduta:2018aqy}. We will use the EoS described in Ref. \cite{Hempel_2010} which is publicly available at the CompOSE database \cite{Typel2015,Oertel:2016bki,Typel:2022lcx}, smoothly matching it to the EoS of the homogeneous core. The small inaccuracies of this treatment can affect the radius of the star (specially of the lightest ones). Secondly, while at T=0 the radius is obtained at the point where the pressure in the star is equal to zero, this procedure is ill defined at finite temperature and another criterium needs to be set. Following the approach described in \cite{Raduta:2020fdn}, we first identify the low density domain of the star where $\log(P)$ and $\log(\epsilon)$ can be considered  linear functions of $\log(\rho_B)$ and then we employ this behavior to extrapolate the EoS at lower densities. This approach guarantees that, for low temperatures and $S/A\lesssim 3$, the radii of the stars is computed with reasonable accuracy (< 20\%). For a more elaborate discussion on these topics, see \cite{Fortin_2016,Raduta:2020fdn}. Furthermore, we note that the results shown here are by no means complete simulations of PN star evolution. Instead, they serve as an indication on how hyperon uncertainties propagate at finite temperature and how these uncertainties would affect the global properties of hot stars. In this sense, although the conditions are inspired by the ones met in PN stars, some of the conclusions can be applied to any phenomena where the matter conditions are similar to the ones described here.

In Fig. \ref{fig:7} we present the temperature and electron fraction profiles employed in the present work. We consider two early-time profiles, $t=1$ s, Fischer (black solid lines) and $t=1$ s, Pascal (purple solid lines) and one intermediate time profile, $t=5.1$ s, Pascal, that assumes a constant entropy $S/A=1$ and constant lepton fraction $Y_e=0.08$ (red lines). In these scenarios, the appearance of muons is suppressed due to the fact that the muonic lepton number is fixed to be $Y_{\mu}=0$. We also consider three profiles of neutrino-free matter at constant entropy values $S/A=1$ (blue lines), $2$ (green lines) and $3$ (yellow lines), representing later evolution stages of the PN star. The electron fractions in these cases are determined by solving the neutrinoless $\beta^{-}$ conditions.

Firstly, we discuss the neutrino-free profiles at constant $S/A$. We can see that the electron fraction on the right panel of Fig.~\ref{fig:7} is correlated with the hyperonic potentials. Softer hyperonic models favour the appearance of hyperons, which in turn lowers the number of leptons, making the deleptonization process more efficient. For larger values of $S/A$, this process starts at lower densities, as the hyperon fraction can then be significantly larger due to the thermal effects. The differences seen in the lepton fraction profiles also reflect in the temperature profiles on the left panel of Fig.~\ref{fig:7}. We observe that a given $S/A$ value can be achieved with lower temperatures in the case of softer models, due the earlier appearance of hyperon species and the corresponding loss of degeneracy.  

\begin{figure}
\centering
\includegraphics[width=0.45\textwidth]{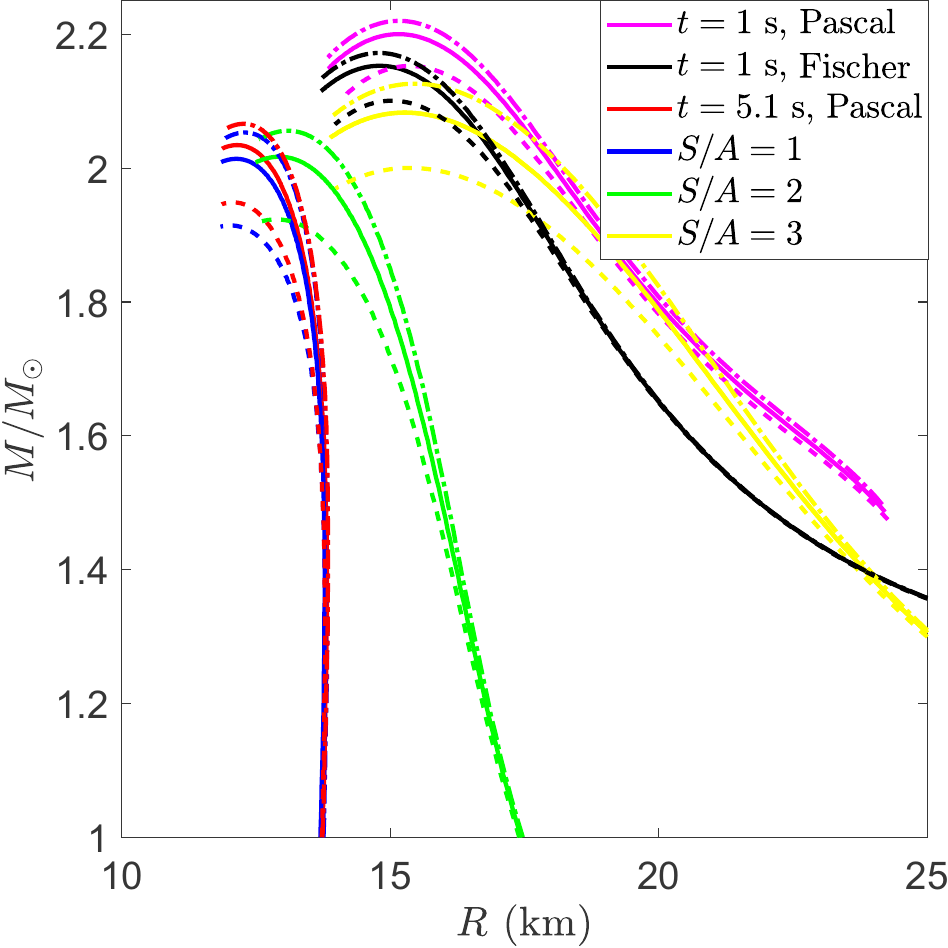}
\caption{$M(R)$ relations for the profiles described in the text. Solid lines represent the results with FSU2H model, dashed lines represent FSU2H$^*$L ones and dashed-dotted lines are results with the FSU2H$^*$U model.}
\label{fig:8}
\end{figure} 

The gravitational mass $M$ is shown in Fig.~\ref{fig:8} as a function of the radius of the star $R$.
We can see that profiles of the star's early evolution, namely ($t=1$ s, Pascal) and ($t=1$, Fischer), favour a smaller compactness, as the stars end up having larger radii and the $M(R)$ curves are the rightmost ones in the figure. One can argue that this is a consequence of the stiffening of the EoS produced by the higher temperatures achieved at the low density part of the core in the early evolution times. We can better illustrate this point by comparing the $M(R)$ curves obtained with the ($t=1$ s, Pascal) (purple curves) and the $S/A = 3$ one (yellow curves) in Fig.~\ref{fig:8}. Although their temperature profiles are similar only up to saturation density, the corresponding $M(R)$ curves are very much alike, regardless of the mass of the star and the stiffness of the hyperonic model that is chosen. This is also strengthened by the observation that, in spite of the temperature profiles of the curves $S/A = 1$ and ($t=1$ s, Pascal) being very similar for densities higher than $\rho_B > 0.3\ \ \text{fm}^{-3}$, their $M(R)$ curves show a completely different behaviour, which is again an indication that it is the EoS at lower densities that governs the compactness of the star. 
It is obvious that the influence of the hyperonic uncertainties can only be seen when hyperons are significantly present in matter, namely when cores can achieve high density and/or high temperatures, so essentially for massive stars ($M > 1.7 M_{\odot}$). The radius of stable, massive ($\sim 2 M_{\odot}$) and hot ($T\gtrsim 50$ MeV) stars predicted by the different hyperonic models can differ by up to 20\%, as can be especially seen upon comparing the different purple or yellow lines in Fig.~\ref{fig:8}.  In similarly cooler scenarios, such as those of the neutrinoless de-leptonized $S/A = 1$ profile and the constant lepton fraction ($t=5.1$ s, Pascal) profile, we find that the $M(R)$ curves of the  de-leptonized, hyperon-richer case (blue lines) lie below those of the constant lepton fraction case (red lines), particularly when hyperons feel less repulsion, as in the FSU2H$^*$L model.
%, the difference is larger, as the fixed lepton fraction hinders the appearance of additional hyperons.

    We can conclude that, as in the $T=0$ case, the uncertainties of the hyperon couplings affect the mass-radius relation only for the most massive stars, where the central density is the highest, and hence the impact of the hyperons is the biggest. These effects are 
   further magnified by temperature, which accentuates the relative differences between the radii of the most massive stars calculated using the FSU2H$^*$L and FSU2H$^*$U models, leading to deviations of up to a few tens of percent.

\begin{figure}
\centering
\includegraphics[width=0.45\textwidth]{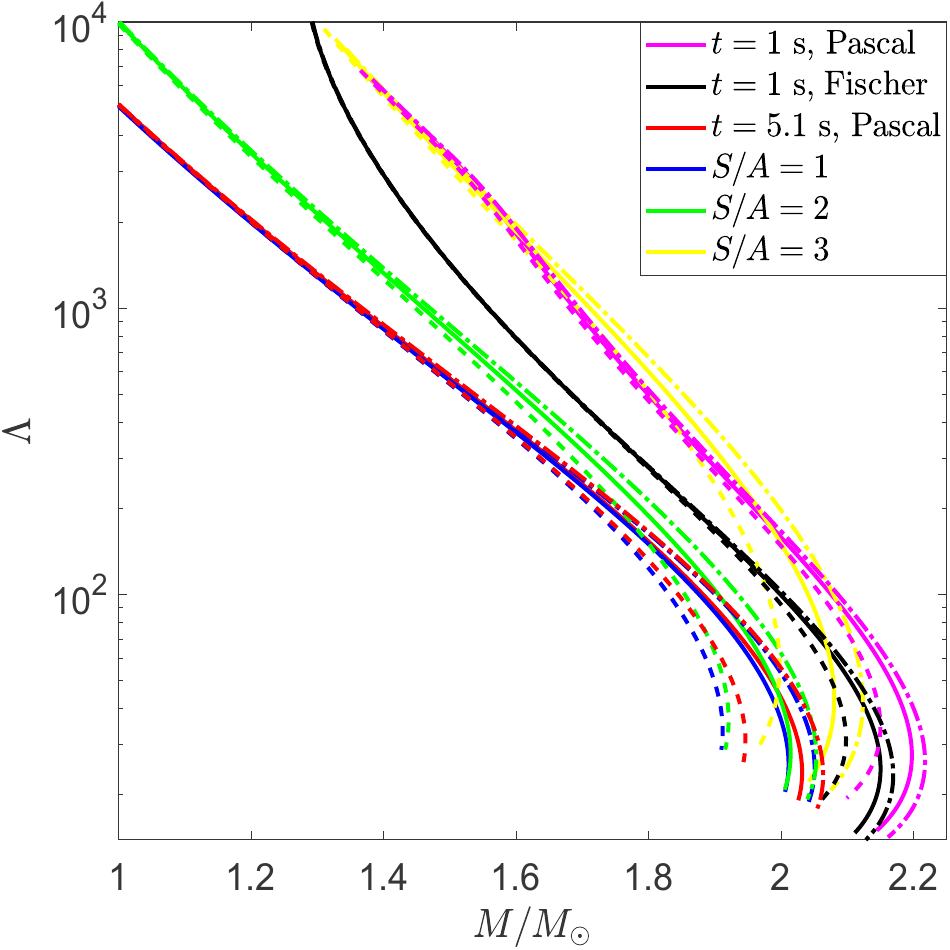}
\caption{$\Lambda(M)$ relations for the profiles described in the text. Solid lines represent the results with FSU2H model, dashed lines represent FSU2H$^*$L ones and dashed-dotted lines are results with the FSU2H$^*$U model.}
\label{fig:9}
\end{figure}

Next, the tidal deformability of hot stars is shown in Fig. \ref{fig:9}. We observe that the behavior of the tidal deformability is also tied to the low density behavior of the temperature profile. Higher values of temperature at low densities produce higher values of the tidal deformability. As for the effect of the hyperonic uncertainties, we observe that softer (stiffer) EoSs, computed with the FSU2H$^*$L (FSU2H$^*$U) model, tend to lower (increase) the tidal deformability, an effect that is more evident for higher mass stars ($M > 1.7 M_{\odot}$).  In the region of most massive stars ($M\gtrsim 2M_{\odot}$) the difference in the predictions of the models for the tidal deformability can be of an order of magnitude.

\begin{figure}
\centering
\includegraphics[width=0.45\textwidth]{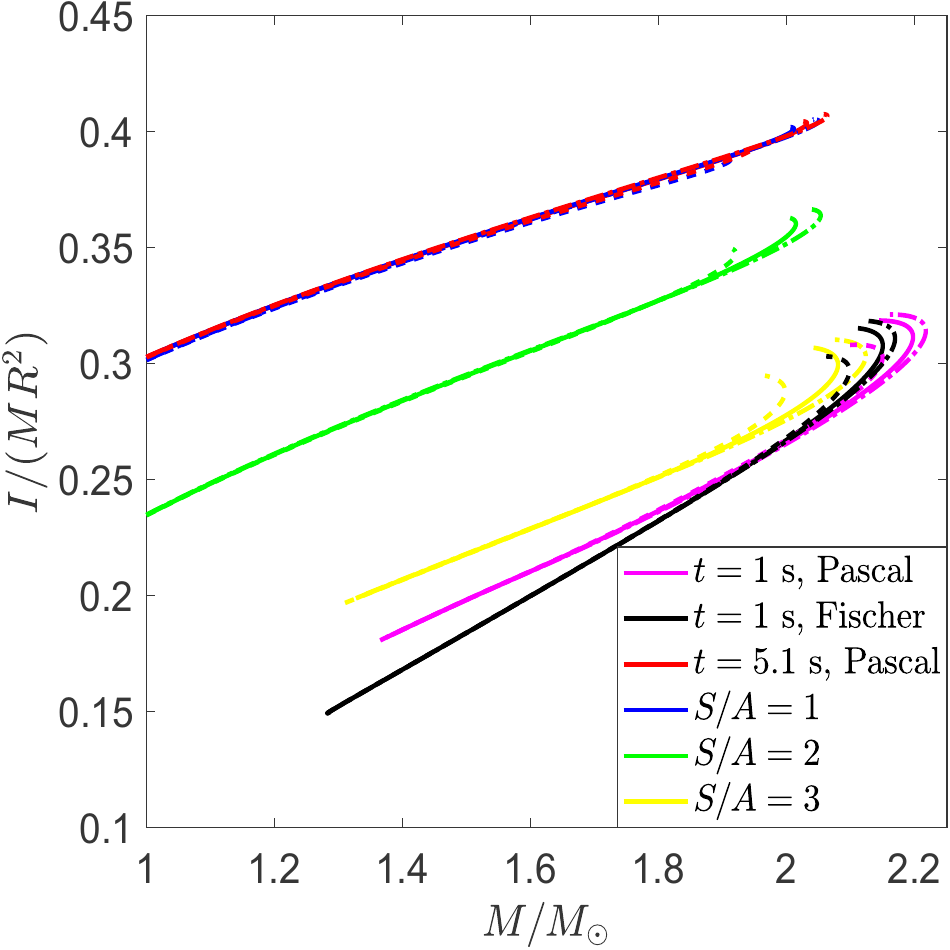}
\caption{$I/MR^2(M)$ relations for the profiles described in the text. Solid lines represent the results with FSU2H model, dashed lines represent FSU2H$^*$L ones and dashed-dotted lines are results with the FSU2H$^*$U model.}
\label{fig:10}
\end{figure}

The dimensionless moment of inertia, $I/(MR^2)$, is shown in Fig. \ref{fig:10} as a function of the mass of the star. One can observe that colder stars tend to have higher dimensionless moment of inertia. This is a direct consequence of their much smaller radii for a given mass. It is noticeable that the dimensionless moment of inertia is not very sensitive to the hyperonic uncertainties, as only small differences can be observed between the predictions of the models.  

\begin{figure}
\centering
\includegraphics[width=0.45\textwidth]{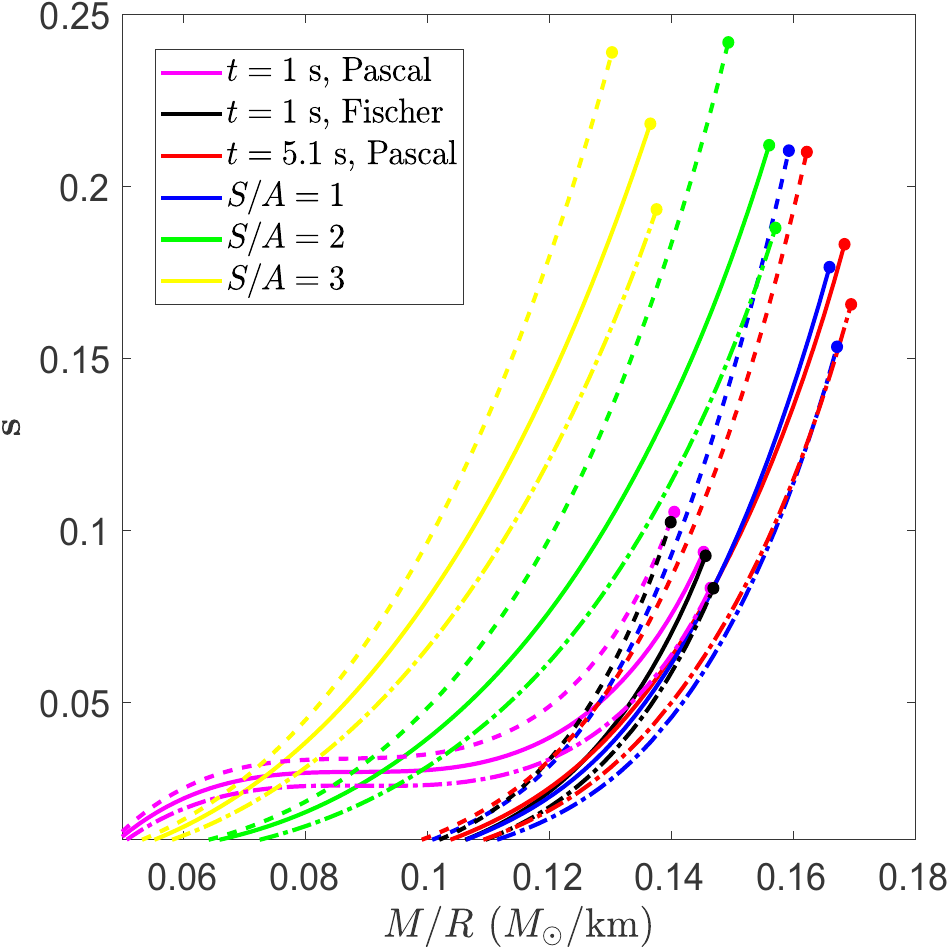}
\caption{Normalized strangeness number $s$ for the profiles described in the text. Solid lines represent the results with the FSU2H model, dashed lines indicate the FSU2H$^*$L ones and dashed-dotted lines are results with the FSU2H$^*$U model. Each line ends with a dot that indicates the maximum mass star configuration that is stable.}
\label{fig:11}
\end{figure}

We finalize by presenting the normalized strangeness number as a function of the compactness of the star for the different temperature profiles explored in this section. The results clearly show that the normalized strangeness number, signalling the presence of hyperons in the star, is very sensitive to the temperature profile in the higher density region. As we can see in the figure, the model $S/A=3$ produces the largest values of $s$ for a given compactness. On the contrary, stars with lower temperatures need higher compactness to produce hyperonic matter. We also note that the profiles with conserved and higher lepton fractions tend to have noticeably lower values of $s$, as the high lepton fraction hinders the appearance of hyperons. Obviously, the normalized strangeness number is also sensitive to the hyperonic uncertainties, being larger (smaller) for the soft FSU2H$^*$L (stiff FSU2H$^*$U) version of the hyperon potentials models.

\section{Conclusions}\label{sec-conclusions}

The main goal of this work is to study the uncertainties of the hyperon potentials  on the finite temperature hyperonic matter EoS in a systematic way. Starting from the FSU2H$^*$ model, we develop two additional models, FSU2H$^*$L and FSU2H$^*$U, obtained by spanning the uncertainty range of the hyperon potentials. These uncertainties have an effect on the composition  and the EoS for different conditions of density, temperature and baryon charge fraction, being especially manifest for ultra dense matter ($\rho_B > 0.7\ \ \text{fm}^{-3}$) when hyperons already achieve a significant abundance. This effect is most noticeable for the pressure, as the differences arising from the hyperonic uncertainties start to be more important than the temperature corrections to the EoS.

Since most of the constraints that we have from astrophysical measurements are obtained from cold NSs, we have first determined the global properties of NSs employing our three hyperonic models in a $T=0$ framework. We have seen that the main effect of the uncertainties on the EoS directly translates into the value of the maximum star mass predicted by the models. Quantitatively speaking, the uncertainty of the maximum mass is around 7\%, while the radii of the maximum mass stars are less affected. However, for a given massive star, the predicted radius greatly differs between the models. As for the tidal deformability and moment of inertia results, we again find the hyperon uncertainties to start playing a role in stars with masses higher than $M>1.5M_{\odot}$, with lower (higher) values of tidal deformabilities and moment of inertia for the softer (stiffer) FSUH$^*$L (FSUH$^*$U) model. The differences are sensibly enhanced in the case of very massive stars ($M> 1.7 M_\odot$).  We note that the information from the gravitational waves on the tidal deformability at this moment constrains the intermediate density region, which is not sensitive to the uncertainties in the hyperonic sector.

Secondly, we have used the FSU2H$^*$L, FSU2H$^*$ and FSU2H$^*$U models to obtain the global properties of stars at finite temperature. While the densities, temperatures and charge fractions explored in the present work are inspired by the typical values found in the evolution of PN stars, we note that the effects discussed here are much more general, and can be applied whenever similar conditions for the EoS are encountered. We find that, due to the increased hyperon abundances at high temperature, the uncertainties in this case play an even bigger role than at zero temperature. For example, the hyperonic uncertainties may produce differences in the radius of the most massive evolved stars of up to 20\%, while the tidal deformability may differ by almost an order of magnitude.

We note that all our results are quite general and not restricted to the specific model used. Thus, similar outcomes are expected from other approaches that take into account hyperonic degrees of freedom. Our findings have a direct implication on relativistic simulations of NS mergers and supernovae, thus emphasizing the need for these simulations to consider hyperons and their uncertainties to ensure the accuracy and reliability of their results.

\section*{Acknowledgements}
We would like to thank Bert Tobias Fischer and Micaela Oertel for providing us the temperature and lepton profiles. We also acknowledge useful discussions with Mario Centelles and Adriana Raduta.

This research was supported from the projects CEX2019-000918-M, CEX2020-001058-M (Unidades de Excelencia ‘María de Maeztu’), PID2019-110165GB-I00, PID2020-118758GB-I00 and PID2022-139427NB-I00, financed by the Spanish Ministry for Science and Innovation - MCIN/ AEI/10.13039/501100011033/FEDER,UE, as well as by the EU STRONG-2020 project, under the program H2020-INFRAIA-2018-1 grant agreement no. 824093. HK acknowledges support from the PRE2020-093558 Doctoral Grant of the Spanish Ministry for Science and Innovation - MCIN/ AEI/10.13039/501100011033/. LT also acknowledges support from the German Research Foundation - DFG through projects No. 315477589 – TRR 211 (Strong-interaction matter under extreme conditions), from the Generalitat Valenciana under contract PROMETEO/2020/023 and from the Generalitat de Catalunya under contract 2021 SGR 171.

%%%%%%%%%%%%%%%%%%%%%%%%%%%%%%%%%%%%%%%%%%%%%%%%%%
\section*{Data Availability}

The data underlying this article will be shared on reasonable request to the corresponding author.



\bibliographystyle{mnras}
\bibliography{example} % if your bibtex file is called example.bib

%\begin{thebibliography}{99}
%\end{thebibliography}

%%%%%%%%%%%%%%%%%%%%%%%%%%%%%%%%%%%%%%%%%%%%%%%%%%

%%%%%%%%%%%%%%%%% APPENDICES %%%%%%%%%%%%%%%%%%%%%

\appendix
\section{Propagation of the hyperonic uncertainties to neutron star observables}
\label{sec:app}

In the main part of the paper we have discussed the results obtained by two extreme model versions, FSU2H$^*$L and FSU2H$^*$U. They were built by making all hyperonic potentials to simultaneously take either the upper or the lower limit of their corresponding uncertainty band. In this appendix we give a proper justification for our choice.

\begin{figure*}
\centering
\begin{subfigure}[t]{0.5\textwidth}
\centering        \includegraphics[height=2.2in]{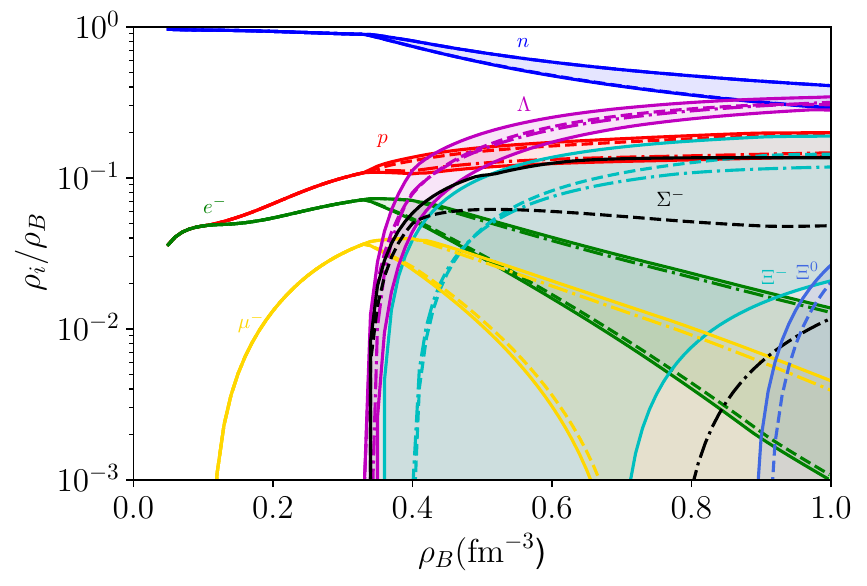}
\caption{$T = 0$}
\end{subfigure}%
\begin{subfigure}[t]{0.5\textwidth}
\centering
\includegraphics[height=2.2in]{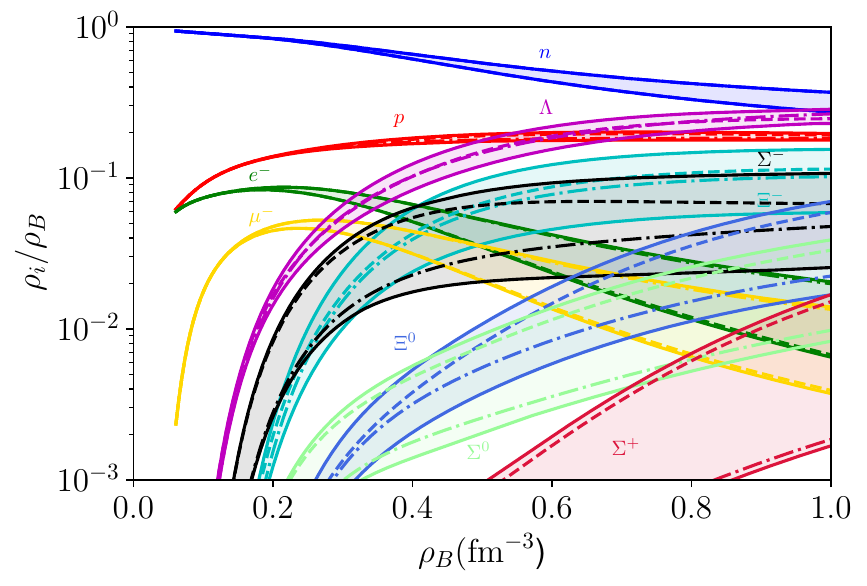}
\caption{$S/A = 2$ }
\end{subfigure}
\caption{Compositions obtained both with the FSU2H$^*$L (dashed lines) and FSU2H$^*$U (dash-dotted lines) parametrizations and the additional ones. For each species, the upper/lower abundance curve signals the density dependence of the maximum/minimum abundance obtained within all the models. Correspondingly, the shaded regions represent the composition uncertainty interval for each species.}
\label{fig:A1}
\end{figure*}

It is clear that the hyperonic uncertainty space is not one-dimensional. This means that different combinations of the softness or hardness of the three hyperonic potentials can in principle affect both the thermodynamical quantities and the observables of the star in a non-linear way. However, as most of the relativistic simulations of neutron star mergers and supernovae are computationally expensive, it would be practical to cover a wide range of the uncertainties with the smallest number of possible models. In order to check whether the two models considered in this work, FSU2H$^*$L and FSU2H$^*$U, can serve this purpose, we have compared their outcome to that of eighteen additional parametrizations. These new models are designed to favor/disfavor the appearance of a particular hyperonic species by taking the smallest/largest value of its corresponding hyperonic potential, while fixing the potentials of the other hyperonic species to their largest/smallest values. This is done for
three diferent values of the $\Lambda \Lambda$ bond energy, namely 0, -0.67, -6.0 MeV, which controls the strength of the hyperon-hyperon interactions. With these eighteen parametrizations we have computed the equation of state, the composition, and both the $M(R)$ and $\Lambda(M)$ relations of cold stars and of stars with constant entropy $S/A = 2$. All stars are assumed to be in $\beta^{-}$ equilibrium. 
%The obtained results from the additional parametrizations are compared with the results of the two extreme models.

The results for the composition of the star are shown in Fig. (\ref{fig:A1}). The solid lines, limiting the corresponding color-shaded regions, stand for the maximum and minimum abundances of each species attained within all parametrizations as functions of the density. Therefore, the shaded regions for each particle can be interpreted as an uncertainty in its abundance. For comparison, the results predicted by the FSU2H$^*$L and FSU2H$^*$U models are represented with dashed and dash-dotted lines, respectively. We can see that the abundance of a particular hyperonic species strongly depends on the interplay between the hyperon uncertainties. The extreme models only partially cover the allowed region. This is especially important for the $\Sigma^-$ and $\Xi^-$ hyperons since, as we discussed in the main part, the uncertainties in their potentials are significant. However, if one focuses on the non-hyperonic species, one observes that the FSU2H$^*$L and FSU2H$^*$U parametrizations do describe their uncertainties very well, both at zero and finite temperature, as the abundance curves predicted by these models lie at the edge of the uncertainty regions. As the total hyperonic content is tied to the non-hyperonic one, one can conclude that the uncertainty of the total number of hyperons in matter is well described with the two extreme models.

\begin{figure*}
\centering
\begin{subfigure}[t]{0.5\textwidth}
\centering        \includegraphics[height=2.2in]{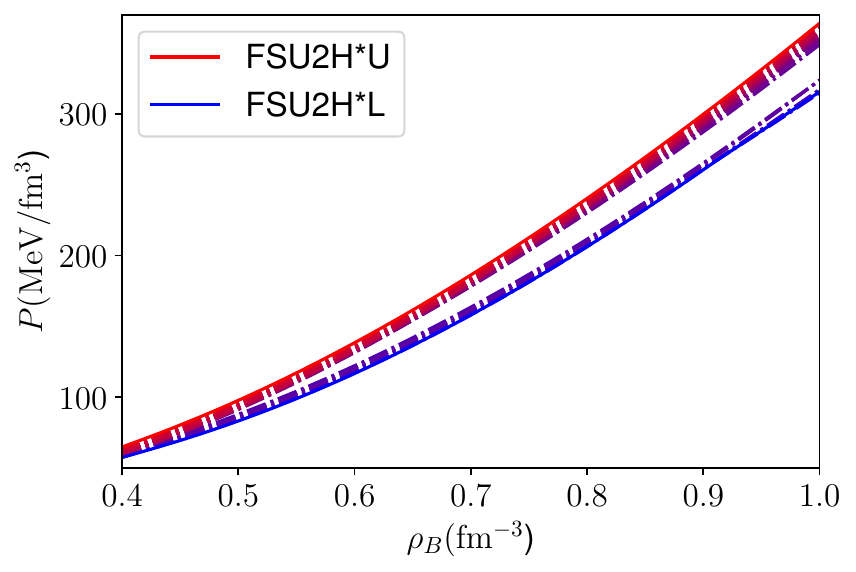}
\caption{$T = 0$}
\end{subfigure}%
\begin{subfigure}[t]{0.5\textwidth}
\centering
\includegraphics[height=2.2in]{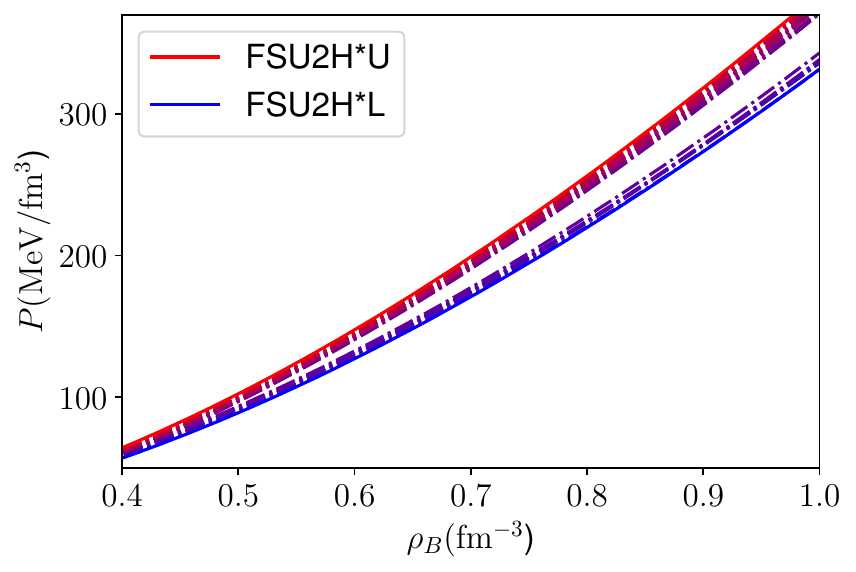}
\caption{$S/A = 2$ }
\end{subfigure}
\caption{$P(\rho_B)$ relation obtained with the original parametrizations, FSU2H$^*$L (blue solid curve) and FSU2H$^*$U (red solid curve), as well as with the additional eighteen parametrizations (dashed lines).}
\label{fig:A2}
\end{figure*}

\begin{figure*}
\centering
\begin{subfigure}[t]{0.5\textwidth}
\centering        \includegraphics[height=2.2in]{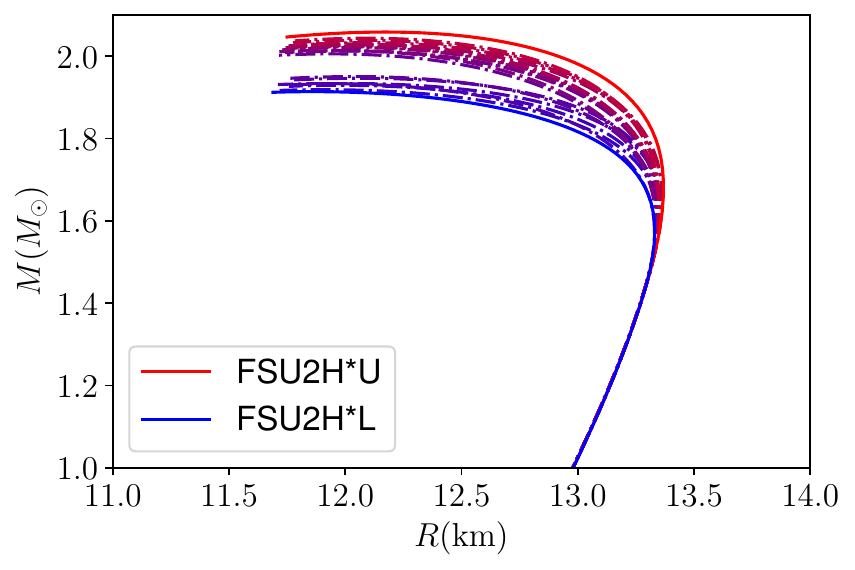}
\caption{$T = 0$}
\end{subfigure}%
\begin{subfigure}[t]{0.5\textwidth}
\centering
\includegraphics[height=2.2in]{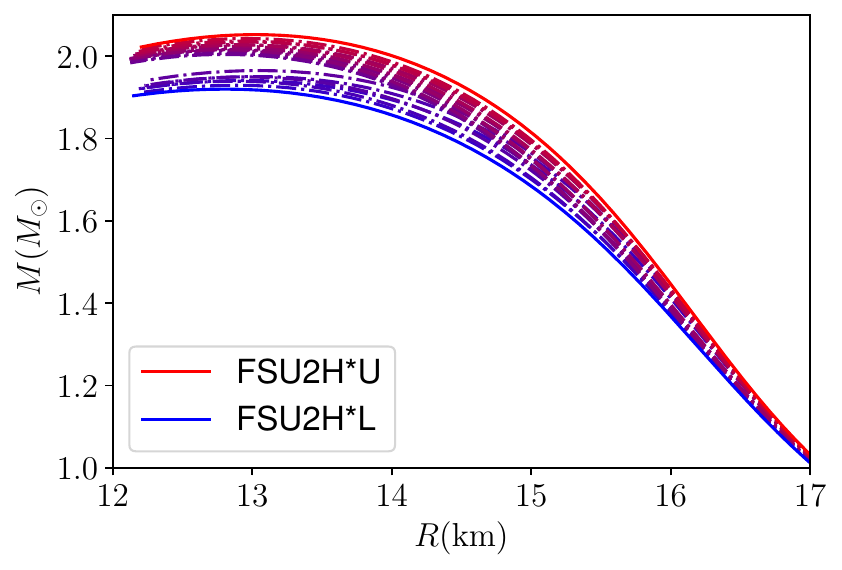}
\caption{$S/A = 2$ }
\end{subfigure}
\caption{$M(R)$ relation obtained with the original parametrizations, FSU2H$^*$L (blue solid curve) and FSU2H$^*$U (red solid curve), as well as with the additional eighteen parametrizations (dashed lines).}
\label{fig:A3}
\end{figure*}

\begin{figure*}
\centering
\begin{subfigure}[t]{0.5\textwidth}
\centering        \includegraphics[height=2.2in]{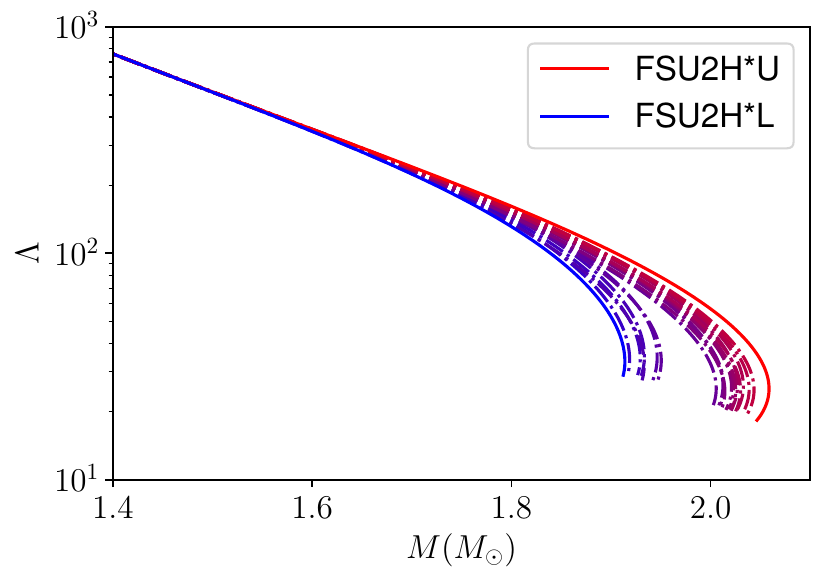}
\caption{$T = 0$}
\end{subfigure}%
\begin{subfigure}[t]{0.5\textwidth}
\centering
\includegraphics[height=2.2in]{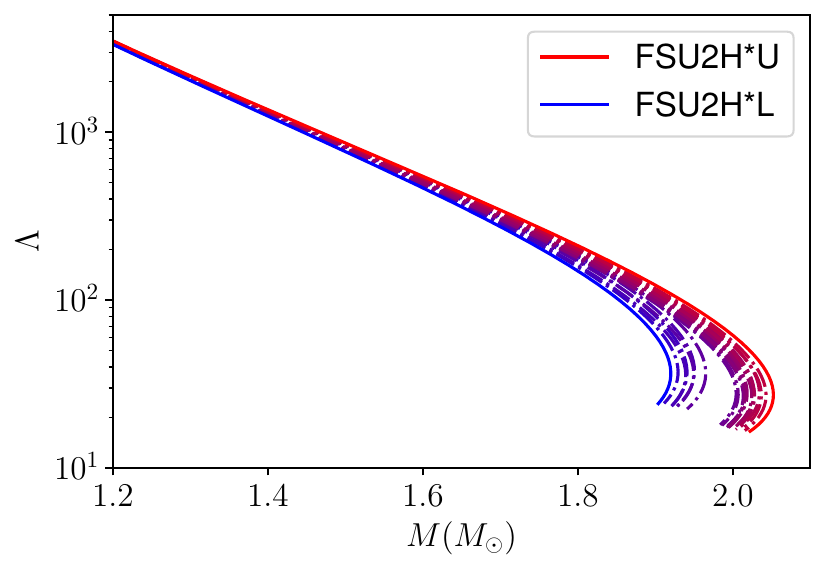}
\caption{$S/A = 2$ }
\end{subfigure}
\caption{$\Lambda(M)$ relation obtained with the original parametrizations, FSU2H$^*$L (blue solid curve) and FSU2H$^*$U (red solid curve), as well as with the additional eighteen parametrizations (dashed lines).}
\label{fig:A4}
\end{figure*}

The conclusion in the main part of the paper about the FSU2H$^*$L and FSU2H$^*$U models describing well the range of uncertainties in the global properties of the star is further confirmed by examining the results of the new models for the thermodynamical quantities and the star observables. In Fig.~(\ref{fig:A2}) we show the pressure of $\beta^{-}$ stable matter as a function of the baryonic density. The dashed lines represent the results of the additional parametrizations, while the solid lines indicate the predictions from the two original models. It is clear that the uncertainty in the pressure of  matter is well described by the FSU2H$^*$U (red solid line) and FSU2H$^*$L (blue solid line) extreme models. The conclusion is valid both at zero and finite temperature. 
This behavior is also found for the star observables, as can be seen in Fig.~(\ref{fig:A3}) and Fig.~(\ref{fig:A4}), which show, respectively, the results for the $M(R)$ relation
and the $\Lambda(M)$ relation of $\beta^{-}$ stable stars obtained by the different parametrizations.

All in all we conclude that the FSU2H$^*$L and FSU2H$^*$U models can be used to explore the uncertainties in the thermodynamical quantities of neutron star matter and the neutron star observables, both at zero and finite temperature. While the uncertainty in the abundance of a particular hyperonic species is only partially described by these two extreme models, they do cover the uncertainty range of the total hyperonic content. Therefore, unless one finds an observable that depends on the specific hyperonic abundances, the FSU2H$^*$L and FSU2H$^*$U models can be safely used in relativistic simulations of neutron star mergers and supernovae to assess the effect of the hyperonic uncertainties in their outcomes.

%%%%%%%%%%%%%%%%%%%%%%%%%%%%%%%%%%%%%%%%%%%%%%%%%%

% Don't change these lines
\bsp	% typesetting comment
\label{lastpage}
\end{document}